\documentclass[iop]{emulateapj}

\usepackage{graphicx}
\usepackage{url}
\usepackage{apjfonts}
\shorttitle{X-ray absorption model for the ISM}
\shortauthors{Gatuzz et al.}

\begin{document}

\title{ISMabs: a comprehensive X-ray absorption model for the interstellar medium}

\author{E.~Gatuzz\altaffilmark{1},
        J.~Garc\'ia\altaffilmark{2},
        T.R.~Kallman\altaffilmark{3},
        C.~Mendoza\altaffilmark{1},
        T.W.~Gorczyca\altaffilmark{4}
        }

\altaffiltext{1}{Centro de F\'isica, Instituto Venezolano de Investigaciones
Cient\'ificas (IVIC), PO Box 20632, Caracas 1020A,  Venezuela
\email{egatuzz@ivic.gob.ve, claudio@ivic.gob.ve}}

\altaffiltext{2}{Harvard-Smithsonian Center for Astrophysics, Cambridge, MA, 02138,  USA
\email{javier@head.cfa.harvard.edu}}

\altaffiltext{3}{NASA Goddard Space Flight Center, Greenbelt, MD 20771, USA
\email{timothy.r.kallman@nasa.gov}}

\altaffiltext{4}{Department of Physics, Western Michigan University, Kalamazoo,
MI 49008, USA \email{thomas.gorczyca@wmich.edu}}

%%%%%%%%%%%%%%%%%%%%%%%%%%%%%%%%%%%%%%%%%%%%%%%%%%%%%%%%%%%%%%%%%%%%%%%%%%%%%%%

\begin{abstract}
  We present an X-ray absorption model for the interstellar medium, to be referred to as {\tt ISMabs}, that takes into account both neutral and ionized species of cosmically abundant elements, and includes the most accurate atomic data available. Using high-resolution spectra from eight X-ray binaries obtained with the {\it Chandra} High Energy Transmission Grating Spectrometer, we proceed to benchmark the atomic data in the model particularly in the neon K-edge region. Compared with previous photoabsorption models, which solely rely on neutral species, the inclusion of ions leads to improved spectral fits. Fit parameters comprise the column densities of abundant contributors that allow direct estimates of the ionization states. {\tt ISMabs} is provided in the appropriate format to be implemented in widely used X-ray spectral fitting packages such as {\sc xspec}, {\sc isis} and {\sc Sherpa}.
\end{abstract}

%%%%%%%%%%%%%%%%%%%%%%%%%%%%%%%%%%%%%%%%%%%%%%%%%%%%%%%%%%%%%%%%%%%%%%%%%%%%%%%


%%%%%%%%%%%%%%%%%%%%%%%%%%%%%%%%%%%%%%%%%%%%%%%%%%%%%%%%%%%%%%%%%%%%%%%%%%%%%%%
\newpage

\section{Introduction}\label{sec_intro}

The physical properties of the interstellar medium (ISM) are coupled to the star life cycles and affect the dynamics of the Galaxy; therefore, the understanding of the ISM composition and phases has broad relevance. The ISM has been studied by a variety of observational techniques, e.g. 21~cm radio emission \citep{dic90}, IR \citep{dwe92}, and UV absorption spectroscopy \citep{jen05}. X-ray high-resolution spectroscopy has lately become an attractive method  as it provides data that are unique and complementary to those gathered with other techniques. By using a bright continuum-dominated X-ray source as a backlight and grating spectrometers, e.g. the {\it Chandra} High Energy Transmission Grating Spectrometer (HETGS) or the {\it XMM-Newton} Reflection Grating Spectrograph (RGS), it is possible through the modeling of the observed absorption features to unravel key ISM physical and chemical conditions. Due to the penetrating power of X-rays through the gas and dust, high-resolution X-ray spectroscopy allows the direct measurement of spectral features such as absorption lines and edges that lead to the identification of multiple atomic ionization species, molecules, solid composites, and to the determination of binding energies \citep{yao09}. Reliable templates are thus needed to diagnose the different absorption signatures in such a diverse and complex plasma.

Useful photoabsorption models to render the ISM features have been progressively refined and incorporated in the X-ray spectral fitting packages. {\tt Wabs}, a model included in the {\sc xspec} X-ray fitting package, was based on a compilation \citep{mor83} of the photoabsorption cross sections by \citet{hen82}, and included the solar abundances of \citet{and82}; however, it neglected contributions from ions, molecules, grains, and abundance enhancements. Due to this latter limitation, \citet{bal92} energy fitted these cross sections for 17 elements in a model referred to as {\tt Phabs}, allowing the treatment of abundances as adjustable parameters. Within the {\sc xspec} spectral fitting package \citep{arn96}, {\tt TBabs} considered improved cross sections and revised abundances as well as contributions from grains and the H$_2$ molecule \citep{wil00}. In order to comply with {\em Chandra} and {\em XMM-Newton} spectral requirements, an updated version of this model ({\tt TBnew}) has been available, which implements finer cross-section meshes to delineate the resonance structures of the oxygen and neon K edges and the iron L edge. It must be noted that all these models only consider photoabsorption by neutral atomic species. On the other hand, \citet{pin10, pin13} have analyzed ISM absorption in low-mass X-ray binary (LMXB) spectra with the {\sc spex} package using three absorption models that assume ionization and energy balance: {\tt hot} for collisionally ionized plasmas; {\tt slab} for plasmas with arbitrary compositions; and {\tt amol} that takes into account the effects of dust and molecules.
\begin{deluxetable*}{lrlcllc}
\tabletypesize{\scriptsize}
\tablecaption{{\it Chandra} HETG Observations \label{tab1}}
\tablewidth{0pt}
\tablehead{
\colhead{Source}  &\colhead{ObsID} & \colhead{Date} & \colhead{Exposure} & \colhead{Read mode} &\colhead{Galactic} & \colhead{Distance} \\
\colhead{ }  &\colhead{ } &\colhead{ } &\colhead{(ks) }&\colhead{ } &\colhead{coordinates}&\colhead{(kpc)}
}
\startdata
4U~1636--53     & 105  & 1999 Oct 20 & 29 & TIMED      & $(332.9, -4.8)$  & $5.95^{a}$ \\
               & 1939 & 2001 Mar 28 & 27 & TIMED      &                  &            \\
               & 6635 & 2006 Mar 22 & 23 & CONTINUOUS &                  &            \\
               & 6636 & 2007 Jul 02 & 26 & CONTINUOUS &                  &            \\
4U~1735--44     &  704 & 2000 Jun 09 & 24 & TIMED      & $(346.0, -6.9)$  & $6.5^{b}$  \\
               & 6637 & 2006 Aug 17 & 25 & CONTINUOUS &                  &            \\
               & 6638 & 2007 Mar 15 & 23 & CONTINUOUS &                  &            \\
4U~1820--30     & 1021 & 2001 Jul 21 & 9.6& TIMED      & $(2.7, -7.9)$    & $7.6^{c}$  \\
               & 1022 & 2001 Sep 12 & 11 & TIMED      &                  &            \\
               & 6633 & 2006 Aug 12 & 25 & CONTINUOUS &                  &            \\
               & 6634 & 2006 Oct 20 & 26 & CONTINUOUS &                  &            \\
               & 7032 & 2006 Nov 05 & 47 & CONTINUOUS &                  &            \\
Cygnus~X--1     & 3407 & 2001 Oct 28 & 17 & CONTINUOUS & $(71.3, 3.0)$    & $7.2^{d}$  \\
               & 3724 & 2002 Jul 30 & 8.8& CONTINUOUS &                  &            \\
Cygnus~X--2     & 8170 & 2007 Aug 25 & 65 & CONTINUOUS & $(87.3, -11.3)$  & $2.14^{e}$ \\
               & 8599 & 2007 Aug 23 & 59 & CONTINUOUS &                  &            \\
               &10881 & 2009 May 12 & 66 & CONTINUOUS &                  &            \\
               & 1102 & 1999 Sep 23 & 28 & TIMED      &                  &            \\
GX~9+9         &  703 & 2000 Aug 22 & 20 & TIMED      & $(8.5, 9.0)$     & $8^{f}$    \\
               &11072 & 2010 Jul 13 & 95 & TIMED      &                  &            \\
Sco~X--1        & 3505 & 2003 Jul 21 & 16 & CONTINUOUS & $(359, 23.7)$    & $2.8^{g}$  \\
XTE~J1817--330 & 6615 & 2006 Feb 13 & 18 & CONTINUOUS & $(359.8, -7.9)$ & $1{-}4^{h}$ \\
               & 6616 & 2006 Feb 24 & 29 & CONTINUOUS &                 &             \\
               & 6617 & 2006 Mar 15 & 47 & CONTINUOUS &                 &             \\
               & 6618 & 2006 May 22 & 51 & CONTINUOUS &                 &             \\
\enddata
\tablecomments{Distances are taken from: $^a$\citet{bra92}; $^b$\citet{gall08}; $^c$\citet{mig04}; $^d$\citet{oro99}; $^e$\citet{mas95}; $^f$\citet{zdz04}; $^g$\citet{Bra99} and $^h$\citet{sal06}.}
\end{deluxetable*}
\citet{gat13a} have carried out a study of ISM oxygen K absorption using {\it Chandra} spectra of the LMXB XTE~J1817--330. Spectral fits were performed with {\sc xspec} using a physical model referred to as {\tt warmabs}, which is based on precalculated tabulations of ionic fractions and level populations from the {\sc xstar} photoionization code to generate synthetic emission and transmission spectra. In this work, a model benchmark of the theoretical atomic data for oxygen ions by \citet{gar05} required wavelength shifts for both the K lines and photoionization cross sections in order to fit the observed spectra accurately. Moreover, an attempt to develop a definitive absorption model for neutral oxygen with a new computed cross section revealed standing discrepancies between the laboratory and observed line positions \citep{gor13}. The {\tt warmabs} model has been subsequently used with the adjusted \ion{O}{1} cross sections of \citet{gar05} and \citet{gor13} to analyze ISM oxygen absorption toward eight LMXBs \citep{gat14a}. Oxygen abundances and ionization parameters along different lines of sight were therein determined, finding only small abundance differences due to the theoretical atomic data, and estimating the $N$({\rm O}~{\sc ii})/$N$({\rm O}) column-density ratio to be around 10\%. In the case of magnesium, new photoabsorption cross sections by \citet{has14} in {\tt warmabs} were used to fit the Mg~K edge in {\it XMM-Newton} spectra of the LMXB GS~1826--238 predicting that most of this element was in ionized form.

The {\tt warmabs} model includes the most accurate atomic data for atomic neutrals and ions, but it is constrained by the assumption of ionization equilibrium.  This may not be practical owing to deviations in the ionizing energy distribution, other ionization mechanisms, or the superposition of multiple components along the line of sight. Hence, we present a model of ISM X-ray absorption, to be referred to as {\tt ISMabs}, that takes into account both neutrals and charged ions allowing their column densities to vary as free parameters. We benchmark {\tt ISMabs} by means of fits to high-resolution spectra from X-ray binary sources obtained with the {\it Chandra} HETGS and comparisons with the {\tt TBnew} and {\tt warmabs} models. This procedure has involved a detailed study of the oxygen and neon K edges as well as the iron L edge. The outline of this paper is as follows. In Section~\ref{sec_dat} the data reduction process is briefly summarized, and in Section~\ref{sec_mod} we describe the selected atomic database and the X-ray absorption model developed to render the ISM absorption features. In Section~\ref{sec_fit_mo} we show fit comparisons of the {\tt TBnew} and {\tt ISMabs} models. Results obtained from the broadband fit are reviewed in Section~\ref{sec_broad}, with particular attention to the neon K- and iron L-edge regions, followed by a detailed discussion of the fits in Section~\ref{sec_results}. Finally, in Appendix~\ref{appendix} we give an account of the importance of ionization balance in abundance determinations.

\section{Observations and data reduction}\label{sec_dat}

The most suitable sources for X-ray ISM studies are bright LMXBs as they do not usually show intrinsic spectral features, and preferably with photon fluxes with high signal-to-noise ratios to enable the measurement of the ISM contribution. To hone the present photoabsorption model, we have selected eight candidates from the catalog of LMXBs \citep{liu07} and the Transmission Grating Catalog and Archive (TGCat)\footnote{http://tgcat.mit.edu/} such that the total number of counts in each HETGS spectrum is greater than $10^{6}$. We have also included spectra from the high-mass X-ray binary Cygnus~X--1 due to its high photon flux, giving a total of 25 observations. Observational details are listed in Table~\ref{tab1}, where it is worth adding that the high-energy resolution of the {\it Chandra} gratings is certainly indispensable in the present modeling developments. The spectra have been reduced with the standard CIAO threads\footnote{http://cxc.harvard.edu/ciao/threads/gspec.html}, the zero-order position being estimated in each case with the {\tt findzo} algorithm\footnote{http://space.mit.edu/cxc/analysis/findzo/}. Spectral fitting was carried out with the {\sc xspec} package (version 12.8.1\footnote{http://heasarc.nasa.gov/xanadu/xspec/}).

\section{ISMabs photoabsorption model}\label{sec_mod}
\begin{deluxetable*}{llll}
\tabletypesize{\scriptsize}
 \tablecaption{Cross-sections included on {\tt ISMabs} \label{tab2}}
\tablewidth{0pt}
\tablehead{
\colhead{Ion} & \colhead{Source}&\colhead{Ion} & \colhead{Source} \\
}
\startdata
H	        &\citet{bet57}$^{a}$		&Mg~{\sc i}, Mg~{\sc ii}	, Mg~{\sc iii} 	&\citet{has14}$^{c}$\\
He~{\sc i}	& Opacity Project$^{b}$	&Si~{\sc i}	&\citet{ver95} \\
He~{\sc ii}	&\citet{bet57}		&Si~{\sc ii}, Si~{\sc iii}&\citet{wit09}$^{d}$ \\	
C~{\sc i}, C~{\sc ii}, C~{\sc iii} 	&\citet{ver95}$^{c}$		&S~{\sc i}	&\citet{ver95} \\
N~{\sc i}, N~{\sc ii},N~{\sc iii}	&\citet{gar09a}$^{d}$		&S~{\sc ii}, S~{\sc iii}	&\citet{wit09} \\
O~{\sc i}	&\citet{gor13}$^{e}$	 		&Ar~{\sc i}	&\citet{ver95} \\
O~{\sc ii}, O~{\sc iii} 	&\citet{gar05}$^{d}$				&Ar~{\sc ii}, Ar~{\sc iii} &\citet{wit11}$^{d}$ \\
Ne~{\sc i}	&\citet{gor00a}$^{f}$ 		&Ca~{\sc i}	&\citet{ver95} \\
Ne~{\sc ii}, Ne~{\sc iii}	&\citet{jue06}$^{g}$			&Ca~{\sc ii}, Ca~{\sc iii}	&\citet{wit11} \\
&								&Fe	        &\citet{kor00}$^{h}$\\
\enddata
\tablenotetext{a}{Analytic method.}
\tablenotetext{b}{Calculated with the $R$-matrix method.}
\tablenotetext{c}{Semiempirical fitting formula based on Hartree--Dirac--Slater computations.}
\tablenotetext{d}{Calculated with the Breit--Pauli $R$-matrix method.}
\tablenotetext{e}{Derived from an analytical formula.}
\tablenotetext{f}{Based on $R$-matrix calculations and an optical potential.}
\tablenotetext{g}{Based on the calculation by \citet{gor00a} and adjusted to astrophysical observations.}
\tablenotetext{h}{Experimental measurements of metallic iron.}
\end{deluxetable*}
ISM photoabsorption causes significant modifications to the observed X-ray spectra, and must therefore be taken into account in any high-resolution spectroscopic analysis. For this purpose, \citet{wil00} developed the {\tt TBabs} and {\tt TBnew} absorption models that have been widely used in the last decade. However, they only include neutral atomic species compromising their ability to render complex ISM spectral structures such as the O, Ne, and Mg K edges \citep{jue04, jue06, has14}. Given this limitation, we have devised a new ISM X-ray absorption model---to be referred to as {\tt ISMabs}---that compiles the most accurate atomic data available for neutrally, singly, and doubly ionized species of H, He, C, N, O, Ne, Mg, Si, S, Ar and Ca; in the case of iron, an experimental cross section has been chosen to model the L-edge absorption region (see Section~\ref{db}).

Photoabsorption in {\tt ISMabs} takes the form
\begin{equation}
I_{\rm obs}(E)=I_{\rm source}(E)\exp(-\tau)
\end{equation}
where $I_{\rm obs}(E)$ is the observed spectrum, $I_{\rm source}(E)$ the spectrum emitted by the source, and $\exp(-\tau)$ is the absorption coefficient. The optical depth $\tau$ is usually defined as
\begin{equation}
\tau = \sigma _{\rm gas}N({\rm H})\ ,
\end{equation}
$\sigma_{\rm gas}$ being the total photoionization cross section comprising all the atoms and $N({\rm H})$ is the hydrogen column density. If a neutral gas is considered, $\sigma _{\rm gas}$ may be written
\begin{equation}
\sigma_{\rm gas}=\sum_{Z}A_{Z}\times\sigma_{Z}
\end{equation}
where $\sigma _{Z}$ is the photoionization cross section of a chemical element with atomic number $Z$, and $A_{Z}$ gives its abundance relative to hydrogen. It is worth noting that, due to the abundance dependency on column density, there is a degeneracy between $A_{Z}$ and $N({\rm H})$, making it possible by varying one or the other to obtain multiple solutions in the calculation of $\tau$. Consequently, we prefer to adopt the following definition for $\tau$:
\begin{equation}
\tau = \sum_{i}^{k}\sigma_{i}(E)N(i)\ ;
\end{equation}
here $\sigma_{i}(E)$ and $N(i)$ are respectively the cross section and column density for the $i$-th ion. We therefore manage the ion column densities, including $N({\rm H})$, as model parameters in order to determine $\tau$, thus avoiding any parametric ambivalence. {\tt ISMabs} always assumes that $N$(\ion{He}{1}) $=N$(\ion{H}{1})$/10$, and as a default, $N$(\ion{He}{2}) $= 0$. But $N$(\ion{He}{2}) is a free parameter which the user can modify if desired. It must be emphasized that there are no changes to the cross sections during fitting except for interpolation; i.e. they are read just once by the code when the model is invoked. The energy range covered by the model is $0.1{-}10$~keV.  The redshift $z$ is also a model parameter, and we will include turbulent broadening in the next revision to {\tt ISMabs} (with the turbulence velocity $v$ as the model parameter in units of km\,s$^{-1}$); however, observations point to a turbulence ($v< 100$~km\,s$^{-1}$) that cannot be detected by {\it Chandra} \citep{gat13a}. Furthermore, for this velocity range, the natural broadening of inner-shell resonances dominate over turbulence broadening.  {\tt ISMabs} has been specially developed for X-ray fitting packages such as {\sc xspec} \citep{arn96}, {\sc isis} \citep{hou00} and {\sc Sherpa} \citep{fre01}.

%%%%%%%%%%%%%%%%%%%%%%%%%%%%%%%%%%%%%%%%%%%%%%%%%%%%%%%%%%%%%%%%%%%%%
\begin{figure*}
  \epsscale{1.0}
  \plotone{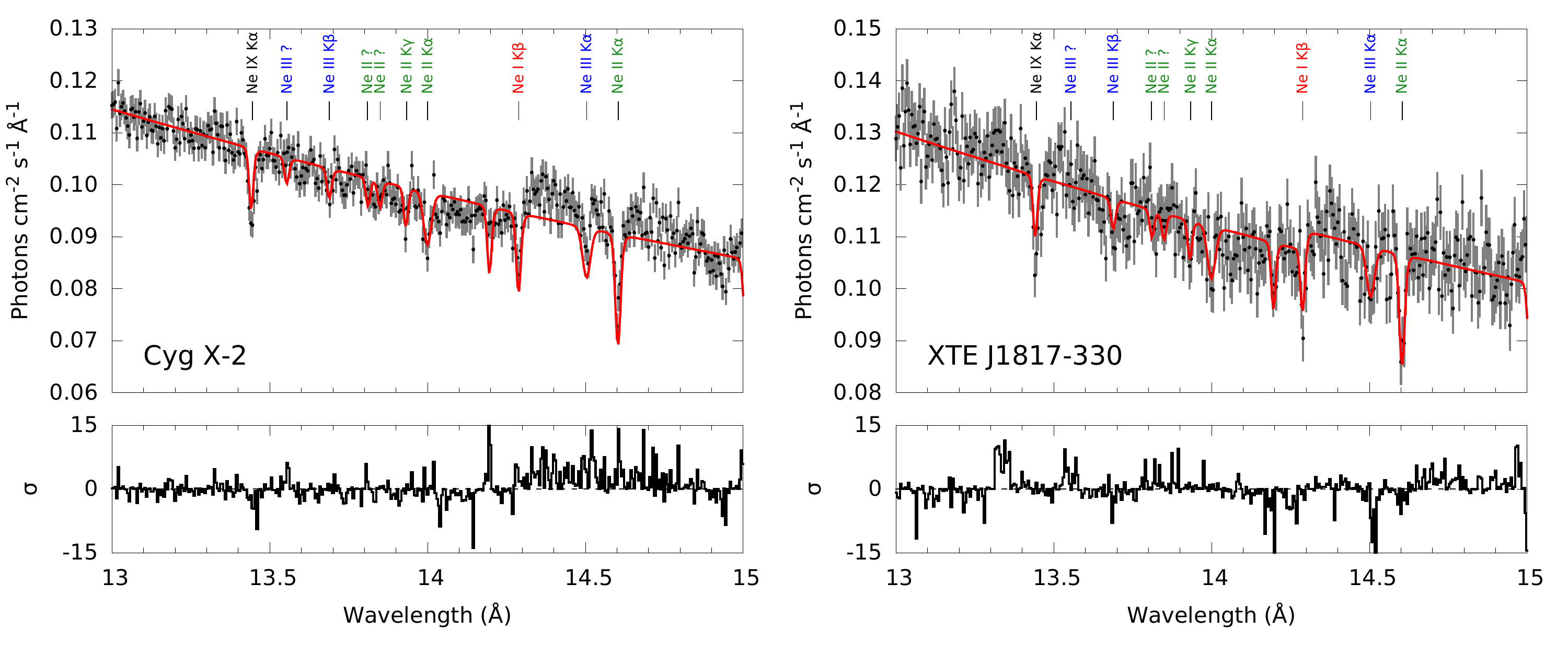}
  \caption{{\it Chandra} MEG flux spectra of the X-ray binaries Cygnus~X--2 (left panel) and XTE~J1817--330 (right panel) simultaneously fitted in the 13--15~\AA\ region using a simple functional model (a power law and several Gaussian profiles). The black data points correspond to the observations while the red solid lines represent the model best fit. The bottom panels show fit residuals in units of $\sigma$. \label{fig1}}
\end{figure*}
%%%%%%%%%%%%%%%%%%%%%%%%%%%%%%%%%%%%%%%%%%%%%%%%%%%%%%%%%%%%%%%%%%%%%

\subsection{Atomic data}\label{sec_atom}
In addition to high signal-to-noise X-ray spectra from a back-light source, reliable atomic data are required to model the ISM absorption signatures. The atomic databases used by many of the currently available X-ray absorption models are not complete and of sufficient accuracy to model grating-resolution spectra. This situation was illustrated in the work of \citet{gat13a}, who used four high-resolution {\it Chandra} spectra of the LMXB XTE~J1817--330 to benchmark the oxygen theoretical atomic database \citep{gar05} in the {\sc xstar} photoionization code. The photoabsorption cross sections of the lowly ionized species required several adjustments: wavelength shifts of the K$\alpha$ resonances in O~{\sc i} ($29$ m\AA) and O~{\sc ii} ($75$ m\AA) and of the whole cross sections for both of these ions. Moreover, a new and more accurate theoretical cross section for O~{\sc i} by  \citet{gor13} showed significant differences in the oscillator-strength distribution, and brought to light a discrepancy in the wavelength scale, namely $33$~m\AA, between the observed  K$\alpha$ line position and the laboratory measurement \citep{sto97}. The quantitative differences in photoabsorption modeling brought about by these two O~{\sc i} cross sections, i.e. \citet{gar05} and \citet{gor13}, were examined by \citet{gat14a} in a study of ISM absorption along several lines of sight, constraining them to less than 5\% in the average ionization parameter and oxygen abundance. A further point of interest is the presence of condensed matter in the ISM, which can be detected through the analysis of X-ray absorption fine structure in the form of spectral modulations near the photoabsorption edges, and by the absence of strong inner-shell resonance transitions below the ionization threshold \citep{lee05,lee09}; however, their proper characterization would require beforehand the mastering of the atomic absorption edges which is the central topic of the present work.

%%%%%%%%%%%%%%%%%%%%%%%%%%%%%%%%%%%%%%%%%%%%%%%%%%%%%%%%%%%%%%%%%%%%%
\begin{figure*}
  \epsscale{1.0}
  \plotone{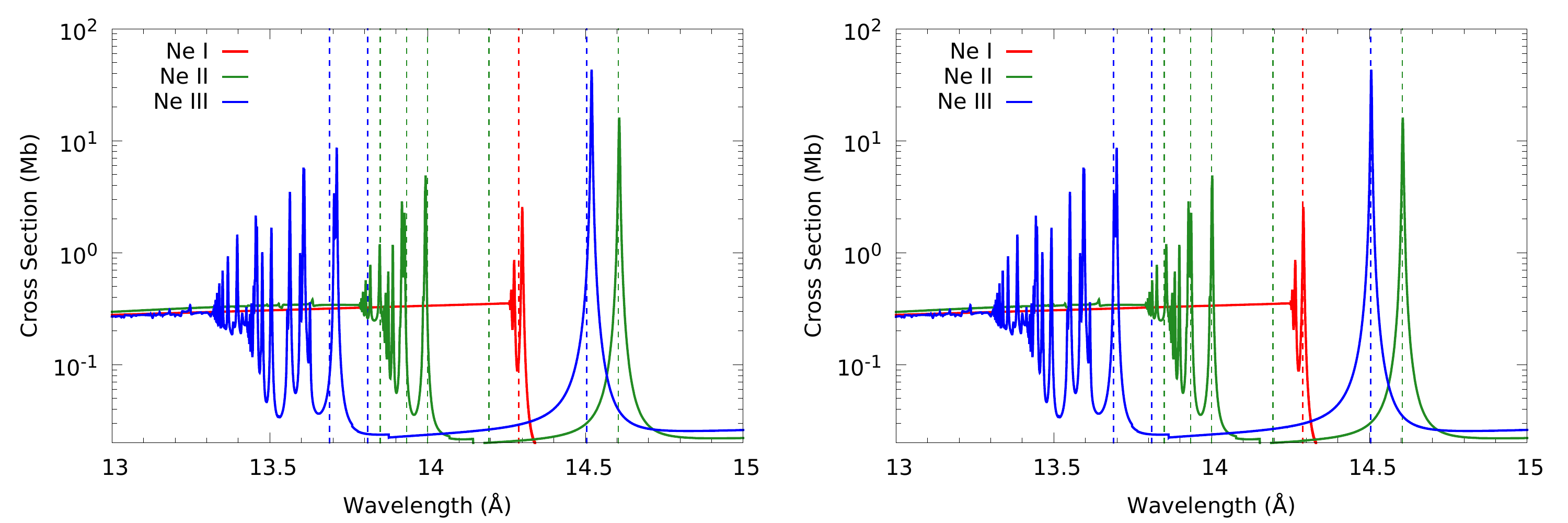}
  \caption{Theoretical photoabsorption cross sections for Ne~{\sc i},  Ne~{\sc ii}, and Ne~{\sc iii} \citep{gor00a, jue06}. The vertical dashed lines are placed at the wavelengths of the observed absorption lines. The left panel displays the original cross sections while the right shows them after the wavelength shifts have been applied. \label{fig2}}
\end{figure*}
%%%%%%%%%%%%%%%%%%%%%%%%%%%%%%%%%%%%%%%%%%%%%%%%%%%%%%%%%%%%%%%%%%%%%

\subsubsection{Database}\label{db}
In Table~\ref{tab2} we list the reference sources for the cross sections incorporated in {\tt ISMabs}. We include the theoretical atomic photoabsorption data for \ion{N}{1}\,--\,\ion{N}{3} by \citet{gar09a}; and by \citet{wit09, wit11} for the singly and doubly ionized species of Si, S, Ar, and Ca, which were calculated with the $R$-matrix method using an optical potential to account for Auger damping. For the neutral species of the latter four elements, we adopt, in the absence of better data, the theoretical cross sections of \citet{ver95} even though they do not take into account the inner-shell resonances that give rise to the K-edge structures. These cross sections are also used for the neutral, singly, and doubly ionized species of C. For oxygen ions, we select the theoretical cross section for \ion{O}{1} by \citet{gor13} and those by \citet{gar05} for \ion{O}{2} and \ion{O}{3}, including the corrections specified by \citet{gat13a} as previously mentioned. In the case of neon, we take the $R$-matrix data (including Auger damping) by \citet{gor00a} for \ion{Ne}{1} and by \citet{jue06} for \ion{Ne}{2} and \ion{Ne}{3}. For the magnesium ions, \ion{Mg}{1}\,--\,\ion{Mg}{3}, we take the photoabsorption cross sections computed by \citet{has14} with the $R$-matrix method that account for orbital relaxation effects to reduce a previous over-estimate of the K-edge position \citep{wit09,wit11}.

%%%%%%%%%%%%%%%%%%%%%%%%%%%%%%%%%%%%%%%%%%%%%%%%%%%%%%%%%%%%%%%%%%%%%
\begin{figure}
  \epsscale{1.0}
  \plotone{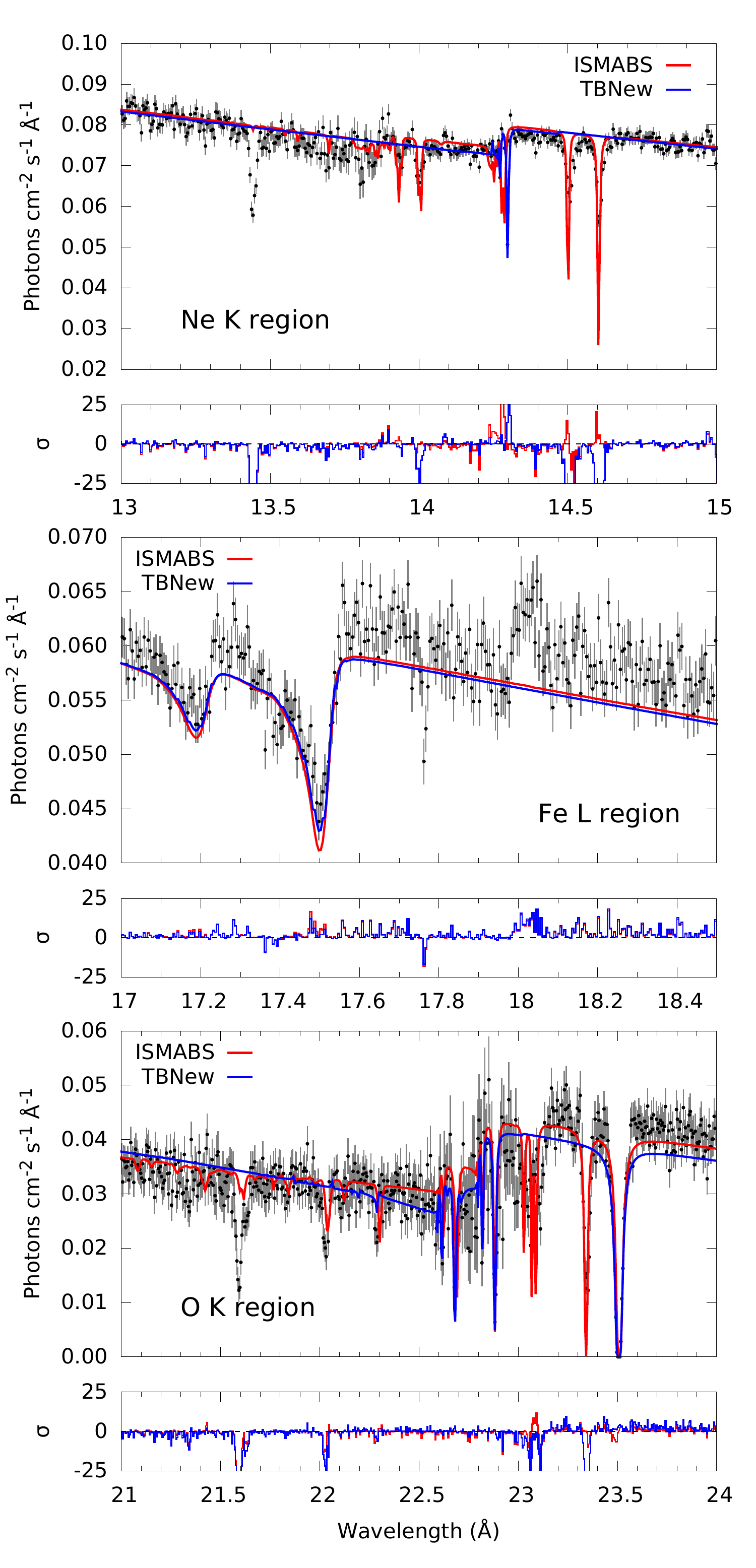}
  \caption{Spectral fits of XTE~J1817--330 using the {\tt ISMabs} and {\tt TBnew} photoabsorption models. The black data points correspond to the observations while the solid lines represent the best fits with {\tt TBnew} (blue line) and {\tt ISMbs} (red line). The bottom panels depict fit residuals in units of $\sigma$. The top, middle, and bottom panels show the Ne K-, Fe L-, and O K-edge regions, respectively. \label{fig3}}
\end{figure}
%%%%%%%%%%%%%%%%%%%%%%%%%%%%%%%%%%%%%%%%%%%%%%%%%%%%%%%%%%%%%%%%%%%%%

In the case of iron, \citet{jue06} fitted the L-edge region using the experimental cross section of metallic iron by \citet{kor00}. On the other hand, \citet{pin10,pin13} adjusted the L~edge using the atomic data for the neutral by \citet{ver95}, but included a molecular solid component since the edge cannot be fitted with a pure gas model. Using {\it Chandra} HETG spectra from five X-ray binaries, \citet{mil09b} adjusted the iron L~edge in the {\tt TBnew} absorption model with the experimental cross section \citep{kor00}. In the present work, we adopt the metallic iron measurements \citep{kor00} but with a wavelength correction of $+39$~m\AA. This shift was originally suggested by \citet{jue06}, who argued that the energy resolution of the laboratory data in combination with the HETG absolute instrumental wavelength accuracy would allow it. We also considered the widely used atomic Fe L-shell cross section of \citet{ver95}, but the resulting spectral fit was inferior to that obtained with the shifted cross section of \citet{kor00}. Finally, it is worth pointing out that both {\tt warmabs} and {\tt ISMabs} use the same atomic database except for Fe since {\tt warmabs} uses the Fe cross-section of \citet{ver95}.

%%%%%%%%%%%%%%%%%%%%%%%%%%%%%%%%%%%%%%%%%%%%%%%%%%%%%%%%%%%%%%%%%%%%%
\begin{deluxetable*}{lclllll}
\tabletypesize{\scriptsize}
 \tablecaption{Ne Absorption Line Assignments\label{tab3}}
\tablewidth{0pt}
\tablehead{
\colhead{Ion} & \colhead{Transition} & \colhead{$\lambda^a$} & \colhead{$\lambda^b$} & \colhead{$\lambda^c$} & \colhead{$\lambda^d$}& \colhead{$\lambda^e$} \\
\colhead{} & \colhead{} & \colhead{(\AA)} & \colhead{(\AA)}	& \colhead{(\AA)} & \colhead{(\AA)} & \colhead{(\AA)}
}
\startdata
Ne~{\sc iii} & K$\gamma$ & 13.453 & $13.441\pm 0.001$ & $13.446 \pm  0.007 $ (Ne~{\sc ix} K$\alpha$) & $13.444\pm 0.001 $ &
    $13.445\pm 0.001$ \\
Ne~{\sc iii} & ?         &        & $13.587\pm 0.007$ &                                              &                    & \\
Ne~{\sc iii} & K$\beta$  & 13.709 & $13.689\pm 0.020$ & $13.695_{-0.002}^{+0.001}$                   &                    &
    $13.690_{-0.001}^{+0.006}$ \\
Ne~{\sc ii}  & ?         &        & $13.810\pm 0.001$ & $13.828\pm 0.003$ (Ne~{\sc vii} K$\alpha$)   &                    & \\
Ne~{\sc ii}  & ?         &        & $13.850\pm 0.001$ & 	                                         &                    & \\
Ne~{\sc ii}  & K$\gamma$ & ?	  & $13.934\pm 0.002$ & $13.936_{-0.003}^{+0.002}$                   &                    & \\
Ne~{\sc ii}  & K$\beta$  & 13.992 & $14.000\pm 0.002$ & $14.004\pm 0.001$                            &                    &
     $14.001_{-0.001}^{+0.002}$ \\
             &           &        & $14.195\pm 0.005$ & $14.212_{-0.029}^{+0.011}$ (Ne~{\sc v} K$\alpha$)&                &\\
Ne~{\sc i}   & K$\beta$  & 14.298 & $14.289\pm 0.001$ & $14.295_{-0.002}^{+0.001}$                       &$14.295\pm 0.003$ &
 	$14.294\pm 0.001$ \\
Ne~{\sc iii} & K$\alpha$ & 14.516 & $14.518\pm 0.001$ & $14.508\pm 0.009$ 	                             &$14.508\pm 0.002$ &
    $14.507\pm 0.002$  \\
Ne~{\sc ii}  & K$\alpha$ & 14.605 & $14.605\pm 0.002$ & $14.608\pm 0.007$                                & $14.608\pm 0.002$ &
    $14.605\pm 0.001$	   \\
\enddata
\tablenotetext{a}{\citet{gor00a} and \citet{jue06}.}
\tablenotetext{b}{Simultaneous functional fits with Gaussian profiles for Cygnus X--2 and XTE J1817--330.}
\tablenotetext{c}{\citet{lia13}.}
\tablenotetext{d}{\citet{jue04}.}
\tablenotetext{e}{\citet{yao09}.}
\end{deluxetable*}
%%%%%%%%%%%%%%%%%%%%%%%%%%%%%%%%%%%%%%%%%%%%%%%%%%%%%%%%%%%%%%%%%%%%%

\subsubsection{Ne benchmark}

Following the benchmark of the oxygen cross sections by \citet{gat13a}, we proceed to compare the theoretical and observed absorption lines in the Ne-edge region of spectra with high signal-to-noise fluxes, namely Cygnus~X--2 and XTE~J1817--330. Figure~\ref{fig1} shows the best fits in the 13--15~\AA\ interval using a simple functional model, i.e. a powerlaw and several Gaussian profiles. For each source, all observations are fitted simultaneously, but the spectra are combined for illustrative purposes; the black data points correspond to the observations while the red solid lines represent the model best fit for each case.  The bottom panels show fit residuals in units of $\sigma$. Several absorption lines are clearly observed including those due to the K$\alpha$ transitions of \ion{Ne}{2} and \ion{Ne}{3}. Identifications of the spectral features and line wavelength positions are listed in Table~\ref{tab3} where the estimates by \citet{jue04}, \citet{yao09}, and \citet{lia13} are also tabulated for comparison. We have identified absorption lines due to transitions in \ion{Ne}{1}, \ion{Ne}{2}, and \ion{Ne}{3}, and the line at $14.195\pm 0.005$ \AA\ presumably belongs to the K$\alpha$ transition of \ion{Ne}{5}. A comparison between the theoretical resonance positions with those determined from the spectral fits reveal the inaccuracies of the theoretical atomic data.

The left panel of Figure~\ref{fig2} shows with solid lines the theoretical cross sections for \ion{Ne}{1}\,--\,\ion{Ne}{3} \citep{gor00a, jue06} that are included in the {\tt ISMabs} model. The vertical dashed lines are placed at the positions of the absorption features determined from the functional fits. In Table~\ref{tab3} we give the line assignments based on a comparison between our functional fits and the theoretical atomic data. Identifications made in previous work using astrophysical data are also listed: Cygnus~X--2 was analyzed by  \citet{jue06} and \citet{yao09} while XTE~J1817--330 was included by \citet{lia13} in their analysis of 11 LMXBs. It may be noted that the \ion{Ne}{3} K$\alpha$ position as well as those with $n>2$ are displaced with respect to the observations, and the cross sections of \ion{Ne}{1} and \ion{Ne}{2} also show similar displacements. In this respect, \citet{jue06} have previously indicated a similar wavelength offset using high-resolution {\it Chandra} spectra. In what follows we have adjusted the K$\alpha$ resonance positions, and shifted the cross sections for the continuum and higher resonances for these three species in order to obtain the best possible agreement with the observed lines. The shifts are, respectively, $-3.2$ m\AA\ and $-15.7$ m\AA\ for the Ne~{\sc ii} and Ne~{\sc iii} K$\alpha$ resonances; and $-11$ m\AA\ , $+7.6$ m\AA, and $-14.7$ m\AA\ for the Ne~{\sc i}, Ne~{\sc ii} and Ne~{\sc iii} cross sections. The right panel of Figure~\ref{fig2} shows the cross sections after the application of the wavelength corrections to be finally included in the {\tt ISMabs} model.

\section{{\tt TBnew} vs. {\tt ISMabs}}\label{sec_fit_mo}

In this section we compare two different X-ray absorption models, namely {\tt TBnew} and {\tt ISMabs}, the former including photoionization cross sections for elements with atomic number $1\leq Z\leq 28$ \citep{wil00} and the contribution from molecules and interstellar grains. It therefore favors the modeling of the observed extinction and depletions in the diffuse ISM. Model parameters are the hydrogen column density ($N({\rm H})$ in units of cm$^{-2}$), elemental abundances ($A_{\rm x}$ relative to solar), and the redshift ($z$). However, {\tt TBnew} only takes into account neutral atomic species, and therefore, the warm or ionized ISM phases cannot be adequately modeled. The {\tt ISMabs} model has been described in detail in Section~\ref{sec_mod}.

%%%%%%%%%%%%%%%%%%%%%%%%%%%%%%%%%%%%%%%%%%%%%%%%%%%%%%%%%%%%%%%%%%%%%
\begin{deluxetable}{lcccccc}
\tabletypesize{\scriptsize}
 \tablecaption{Model Fits For Neutrals \label{tab4}}
\tablewidth{0pt}
\tablehead{
\colhead{Parameter} & \colhead{{\tt TBnew}} & \colhead{{\tt ISMabs}}\\
}
\startdata
$N({\rm H})$          & $1.28 \pm 0.16$ & $1.48\pm 0.31$ \\
$N({\rm O~{\rm I}})$  & $5.38\pm 3.60$  & $8.65\pm 3.21$ \\
$N({\rm Ne~{\rm I}})$ & $1.56\pm 0.42$  & $1.63\pm 0.14$ \\
$N({\rm Fe~{\rm I}})$  & $0.32\pm 0.22$  & $0.54\pm 0.24$ \\
Reduced $\chi^2$      & $2.07$          & $1.78$         \\
\enddata
\tablecomments{$N({\rm H})$ in units of $10^{21}$~cm$^{-2}$. Metal column densities in units of $10^{17}$~cm$^{-2}$.}
\end{deluxetable}
%%%%%%%%%%%%%%%%%%%%%%%%%%%%%%%%%%%%%%%%%%%%%%%%%%%%%%%%%%%%%%%%%%%%%

Figure \ref{fig3} shows the Ne K-, Fe L-, and O K-edge regions in the broadband fit of XTE~J1817--330 with both {\tt TBnew} and {\tt ISMabs}. In each panel, the black data points correspond to the observations while the solid lines represent the best fits with {\tt TBnew} (blue line) and {\tt ISMabs} (red line). The bottom panels depict fit residuals in units of $\sigma$. It is clear that the {\tt ISMabs} residuals are smaller particulary around the neon and oxygen edges. In the Fe L-edge region the residuals are similar since both models use the same experimental atomic data, although small differences are noted perhaps due to the continuum. It is found that the inclusion of charged ions in addition to neutrals leads to a significant improvement in the fits. The difference in the statistics is of $\Delta \chi^2 = 12704$ if the fit is performed with {\tt ISMabs}, with 4 new free parameters. Moreover, as shown in Table~\ref{tab4}, if only the column densities for neutrals are considered (i.e. considering the same number of free parameters), both models give similar results, however the {\tt ISMabs} model performs a better fit, with $\Delta \chi^2 =1559$, due to the higher accuracy in the atomic data included.
 
\section{Broadband Fit}\label{sec_broad}
Following \citet{gat13a}, we have performed a broadband fit (11--24 \AA) of all the spectra listed in Table~\ref{tab1} using {\tt ISMabs} and a power-law model. We rebin the data to at least 20 counts per channel in order to use $\chi^2$ statistics. For each source, all observations are fitted simultaneously varying the power-law normalization and photon index, thus accounting for possible continuum variations between observations that are always detected. The H, O, Ne, and Fe column densities, as well as those of O and Ne ions, are taken as free parameters in {\tt ISMabs}.

%%%%%%%%%%%%%%%%%%%%%%%%%%%%%%%%%%%%%%%%%%%%%%%%%%%%%%%%%%%%%%%%%%%%%
\begin{figure*}
  \epsscale{1.0}
  \plotone{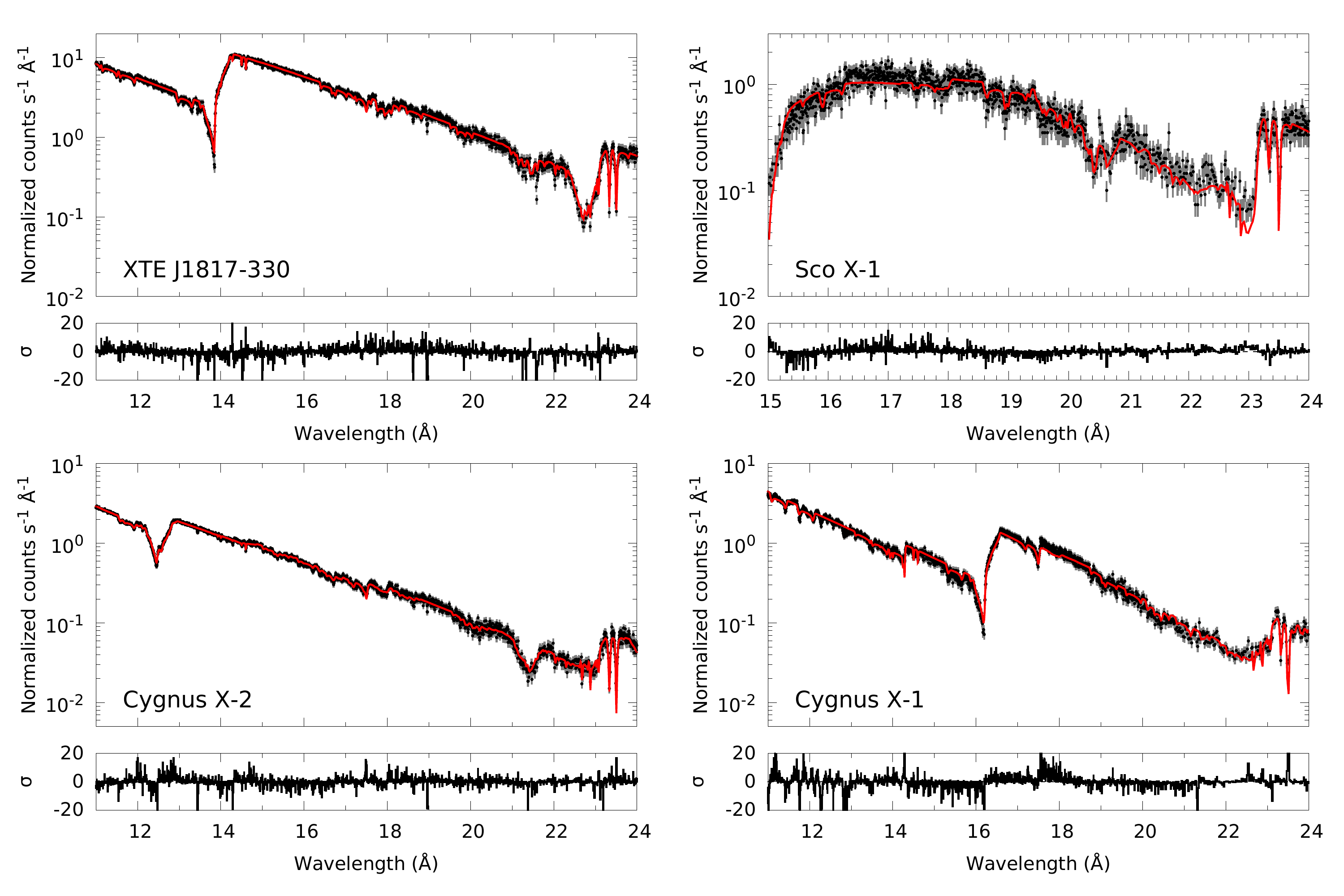}
  \caption{{\tt ISMabs} best broadband fits (solid red lines) for each X-ray source included in the present study. For each source the observations have been combined for illustrative purposes.\label{fig4}}
\end{figure*}
%%%%%%%%%%%%%%%%%%%%%%%%%%%%%%%%%%%%%%%%%%%%%%%%%%%%%%%%%%%%%%%%%%%%%

%%%%%%%%%%%%%%%%%%%%%%%%%%%%%%%%%%%%%%%%%%%%%%%%%%%%%%%%%%%%%%%%%%%%%
\begin{figure*}
  \epsscale{1.0}
  \plotone{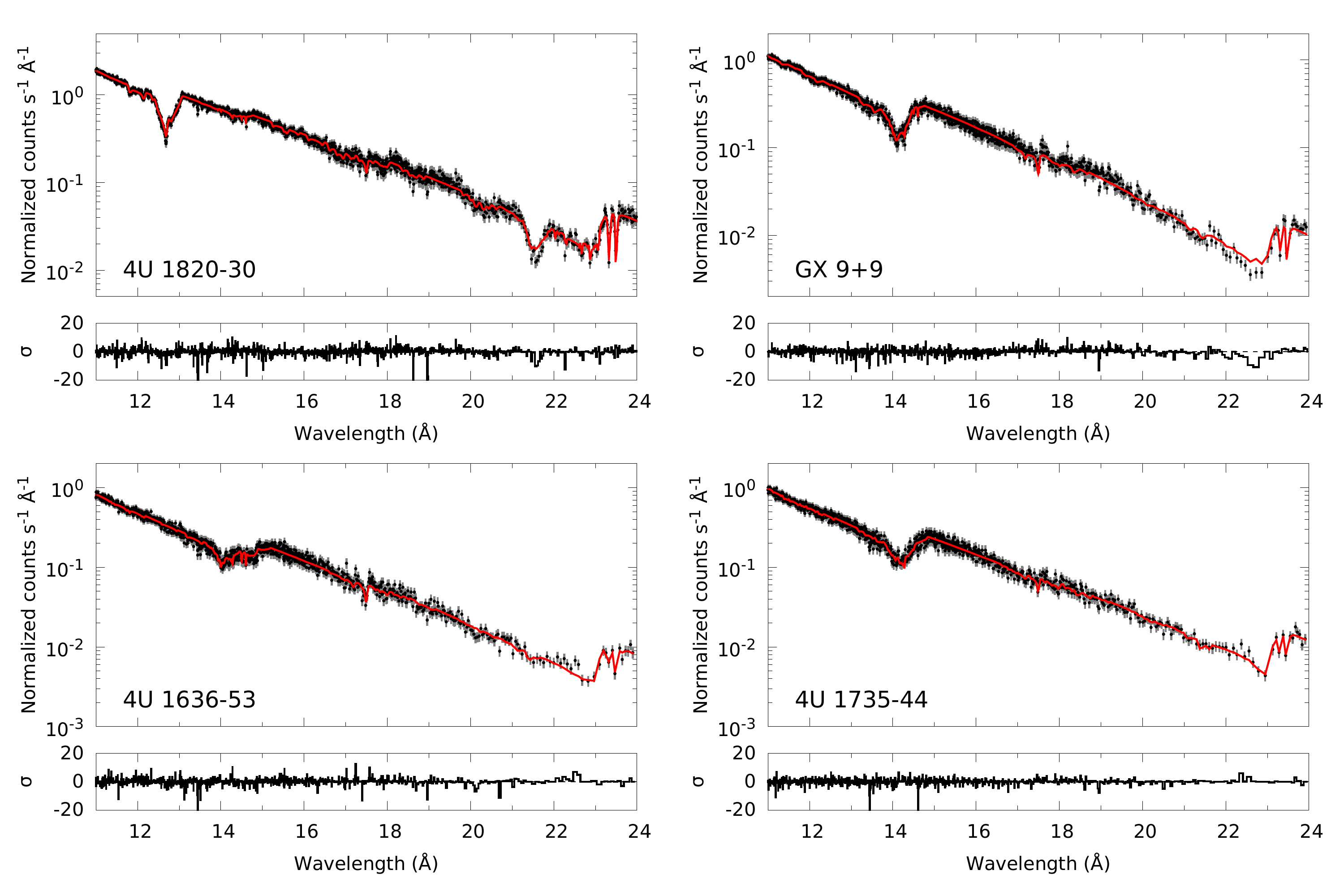}
  \caption{{\tt ISMabs} best broadband fits (solid red lines) for each X-ray source included in the present study. For each source the observations have been combined for illustrative purposes.\label{fig5}}
\end{figure*}
%%%%%%%%%%%%%%%%%%%%%%%%%%%%%%%%%%%%%%%%%%%%%%%%%%%%%%%%%%%%%%%%%%%%%

%%%%%%%%%%%%%%%%%%%%%%%%%%%%%%%%%%%%%%%%%%%%%%%%%%%%%%%%%%%%%%%%%%%%%
\begin{figure*}
  \epsscale{1.0}
  \plotone{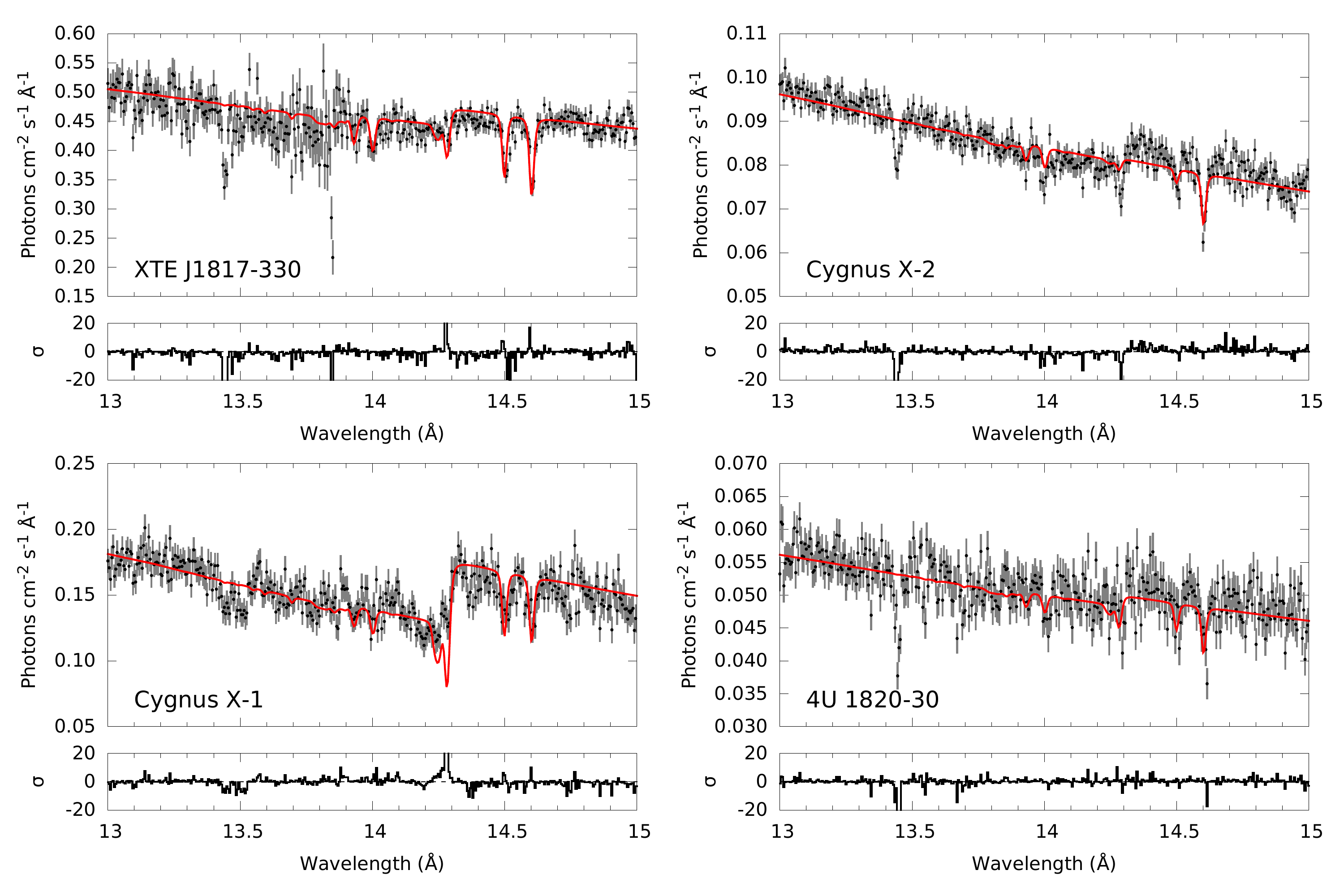}
  \caption{{\tt ISMabs} best broadband fits (solid red lines) in the Ne K-edge region. For each source the observations have been combined for illustrative purposes.\label{fig6}}
\end{figure*}
%%%%%%%%%%%%%%%%%%%%%%%%%%%%%%%%%%%%%%%%%%%%%%%%%%%%%%%%%%%%%%%%%%%%%

%%%%%%%%%%%%%%%%%%%%%%%%%%%%%%%%%%%%%%%%%%%%%%%%%%%%%%%%%%%%%%%%%%%%%
\begin{figure*}
  \epsscale{1.0}
  \plotone{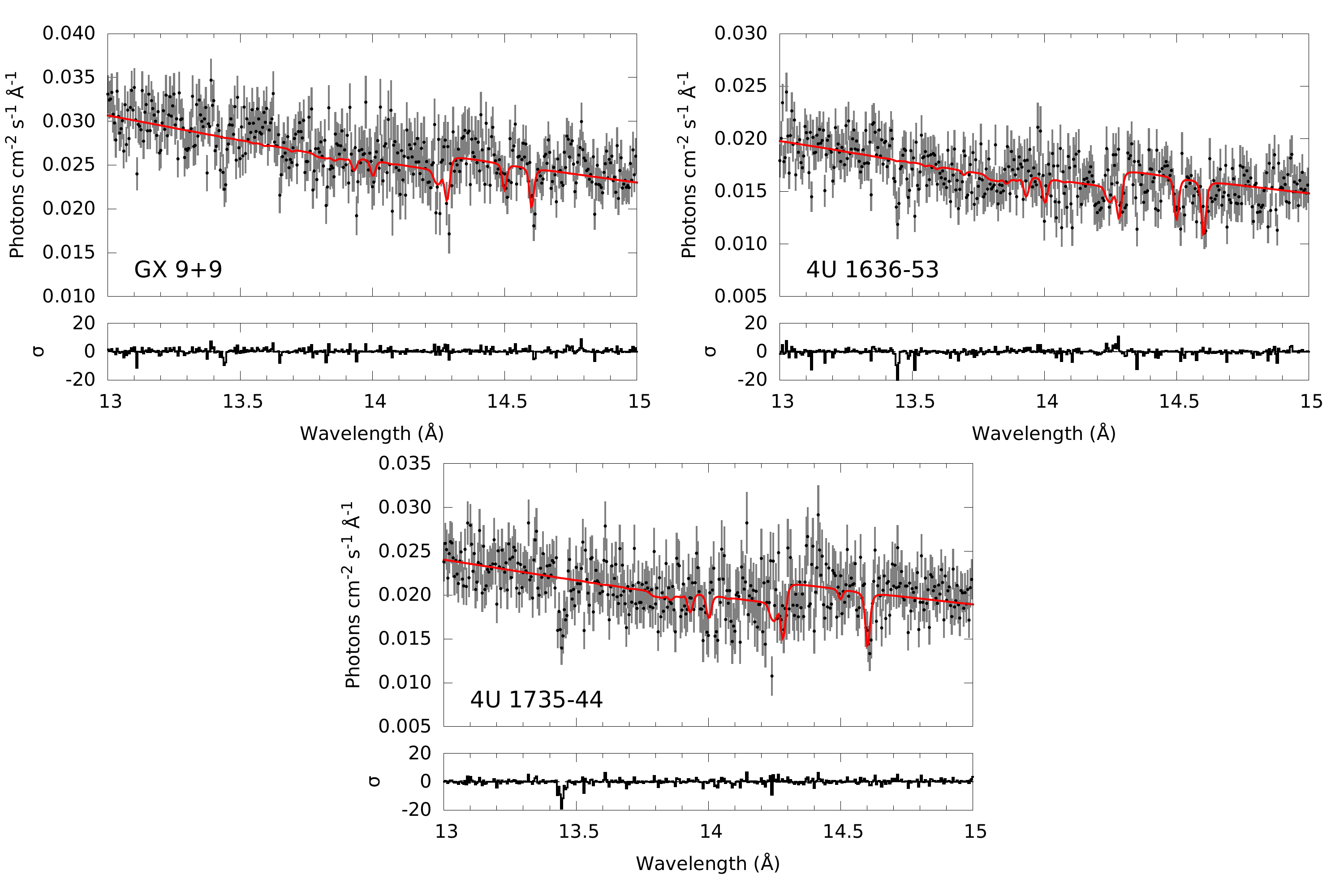}
  \caption{{\tt ISMabs} best broadband fits (solid red lines) in the Ne K-edge region. For each source the observations have been combined for illustrative purposes.\label{fig7}}
\end{figure*}
%%%%%%%%%%%%%%%%%%%%%%%%%%%%%%%%%%%%%%%%%%%%%%%%%%%%%%%%%%%%%%%%%%%%%

%%%%%%%%%%%%%%%%%%%%%%%%%%%%%%%%%%%%%%%%%%%%%%%%%%%%%%%%%%%%%%%%%%%%%
\begin{figure*}
  \epsscale{1.0}
  \plotone{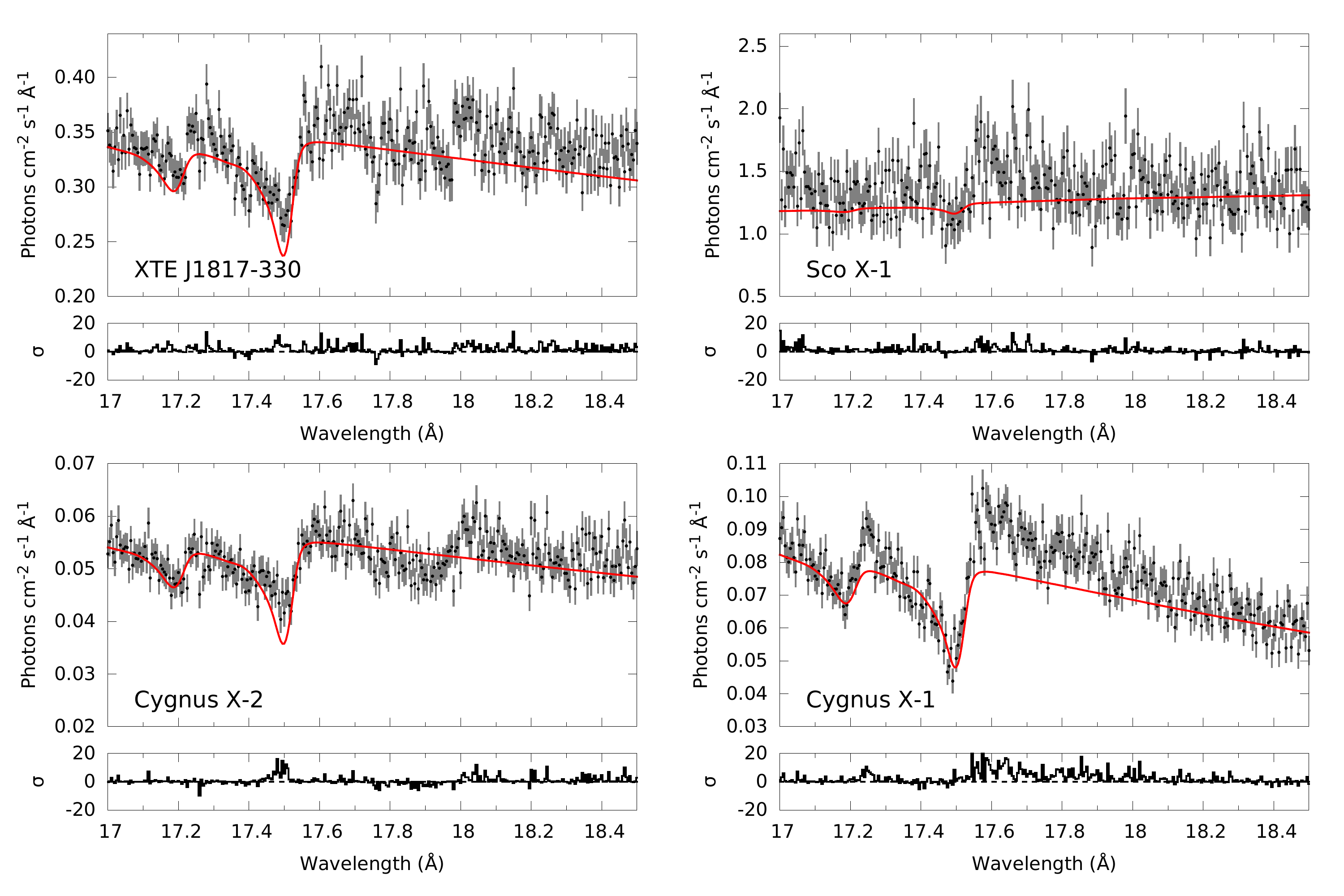}
  \caption{{\tt ISMabs} best broadband fit (solid red lines) in the Fe L-edge region. For each source the observations have been combined for illustrative purposes.\label{fig8}}
\end{figure*}
%%%%%%%%%%%%%%%%%%%%%%%%%%%%%%%%%%%%%%%%%%%%%%%%%%%%%%%%%%%%%%%%%%%%%

%%%%%%%%%%%%%%%%%%%%%%%%%%%%%%%%%%%%%%%%%%%%%%%%%%%%%%%%%%%%%%%%%%%%%
\begin{figure*}
  \epsscale{1.0}
  \plotone{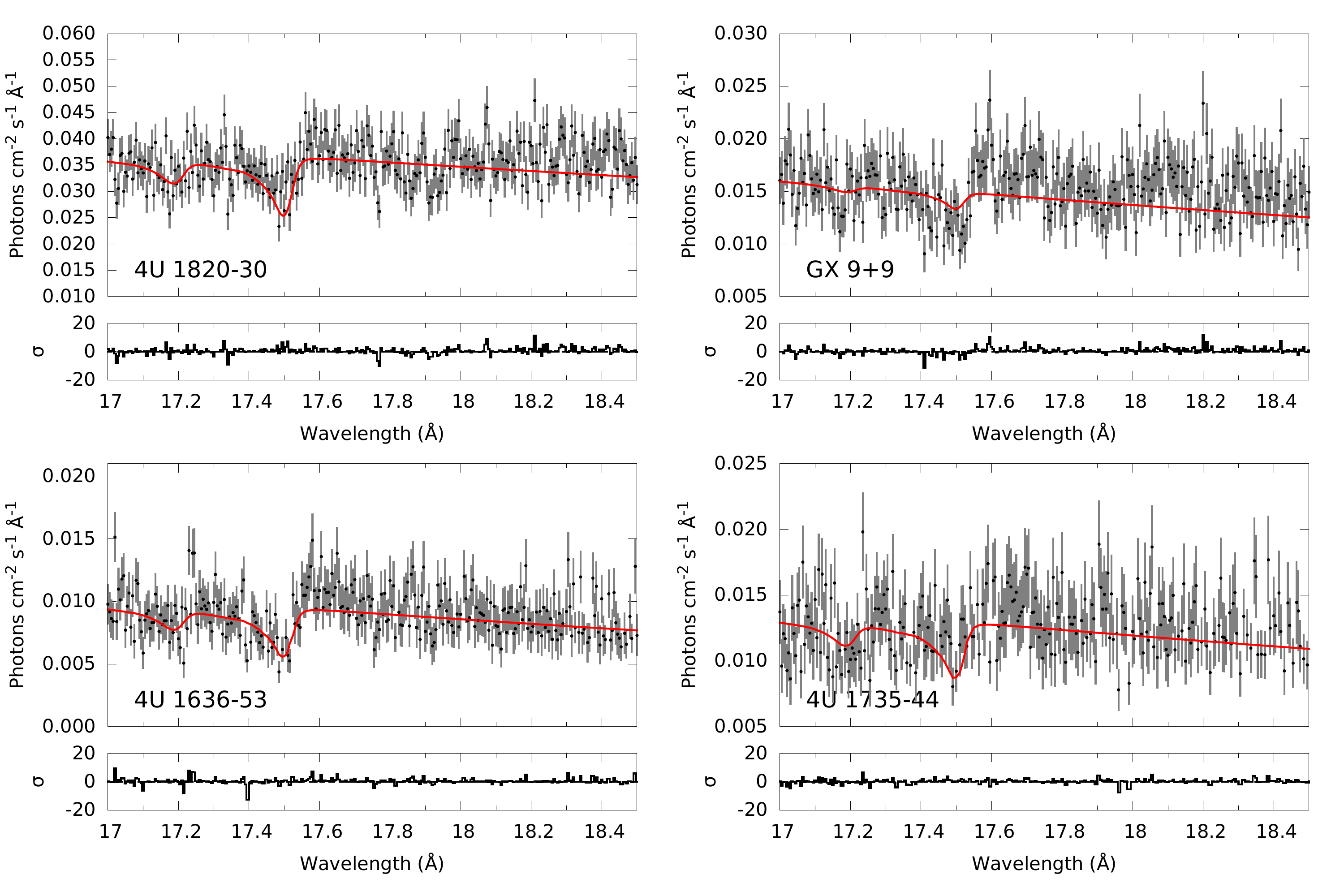}
  \caption{{\tt ISMabs} best broadband fit (solid red lines) in the Fe L-edge region. For each source the observations have been combined for illustrative purposes.\label{fig9}}
\end{figure*}
%%%%%%%%%%%%%%%%%%%%%%%%%%%%%%%%%%%%%%%%%%%%%%%%%%%%%%%%%%%%%%%%%%%%%

Figures~\ref{fig4} and~\ref{fig5} show the best {\tt ISMabs} fits in the  11--24~\AA\ interval. In the case of Sco~X--1 the fit was performed in the 15--24~\AA\ interval as there are no counts below ${\sim}15$~\AA. For each source the observations have been combined for illustrative purposes. It must be noted that, even though this figure looks similar to Figure~1 of \citet{gat14a}, it is now obtained with our new photoabsorption model. Table~\ref{tab5} lists the fit results where it may be appreciated that {\tt ISMabs} yields the ion column densities directly, and a statistically acceptable fit is obtained for each source.

A comparison of the present hydrogen column densities with previous results is given in Table~\ref{tab6} finding satisfactory agreement, i.e. within the error bars. As mentioned in \citet{gat14a}, the column-density variations around Cygnus~X--1 are not resolved in the 21~cm survey; therefore, the values derived by means of spectral fits should be more reliable.

A zoom of the broadband fit in the interval 13--15~\AA\ is depicted in Figures~\ref{fig6} and~\ref{fig7} (Sco~X--1 is excluded for lack of counts below ${\sim}15$~\AA). It may be seen that the absorption features in this wavelength region are more complex than a simple edge: K$\alpha$ absorption lines of \ion{Ne}{2} and \ion{Ne}{2} are detected at $14.60$ \AA\ and $14.51$ \AA, respectively, as well as K$\beta$ from  \ion{Ne}{1}, \ion{Ne}{2}, and \ion{Ne}{3} ($14.30$ \AA, $14.00$ \AA, and $13.68$ \AA, respectively). The line at $13.45$ \AA, attributed to \ion{Ne}{3} K$\gamma$, is observed in all sources. Therefore, the occurrence of such lines justifies the inclusion of accurate cross sections for \ion{Ne}{2} and \ion{Ne}{3} in {\tt ISMabs} in order to procure reliable spectral fits; for instance, in Figures~\ref{fig6} and~\ref{fig7} low residuals are found around the \ion{Ne}{2} K$\beta$ and K$\gamma$ lines at $14.00$ \AA\ and $13.93$ \AA, respectively. Furthermore, a similar zoom is shown in Figures~\ref{fig8} and~\ref{fig9} of the Fe~L edge (17--18.5~\AA) which benchmarks the \ion{Fe}{1} cross section adopted in {\tt ISMabs} (see Section~\ref{db}). It is worth recalling that the selected cross section for \ion{Fe}{1} was obtained from measurements of solid iron, and it seems to fit the observations adequately. This finding supports previous assertions that most of the Fe in the ISM is in solid form \citep{lee09}.
\section{Discussion and Conclusions}\label{sec_results}
\begin{figure}
  \epsscale{1.0}
  \plotone{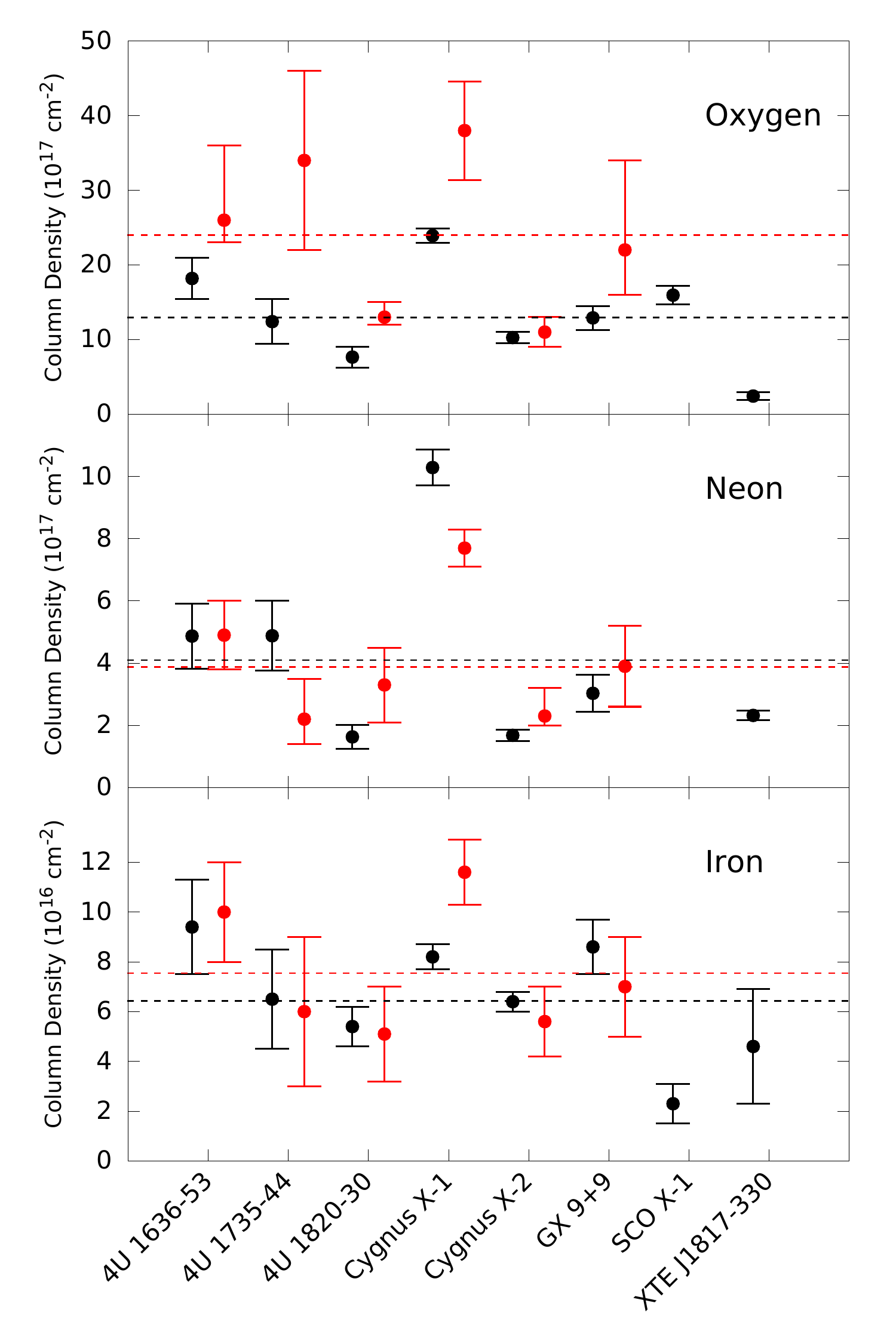}
  \caption{Comparison of the oxygen, neon and iron total column densities obtained from the {\tt ISMabs} model (black lines) and those derived by \citet{jue04,jue06} (red lines). The average ratios are plotted with a dashed horizontal line. \label{fig10}}
\end{figure}

We have developed a new spectral model, referred to as {\tt ISMabs}, to fit ISM X-ray absorption that takes into account photoabsorption cross sections for neutrally, singly, and doubly ionized species of cosmically abundant elements, namely H, He, C, N, O, Ne, Mg, Si, S, Ar, Ca, and Fe. In other words, {\tt ISMabs} enables the determination of the ion column densities directly. Although the predominant component of the ISM is the neutral gas, the inclusion of singly and doubly ionized species provides improved fits when compared with a model (e.g. {\tt TBnew}) that only accounts for neutral absorption. This shows that the understanding of the relevant atomic systems and processes, as well as the accuracy and completeness of the relevant atomic data, are both crucial to avoid incorrect interpretations of the absorption structures detected in X-ray high resolution spectra. This may occur in studies of ISM X-ray absorption that rely on fits with Gaussian profiles.
%%%%%%%%%%%%%%%%%%%%%%%%%%%%%%%%%%%%%%%%%%%%%%%%%%%%%%%%%%%%%%%%%%%%%
\begin{deluxetable}{llllll}
\tabletypesize{\scriptsize}
 \tablecaption{H Column Densities ($10^{21}$~cm$^{-2}$) \label{tab6}}
\tablewidth{0pt}
\tablehead{
\colhead{Source}& \colhead{{\tt ISMabs}} & \colhead{$21$~cm$^a$} & \colhead{Juett$^{b}$} & \colhead{Others}
}
\startdata
4U~1636--53     & $ 4.16\pm 0.42$ & 3.30 & $5.3^{+2.1}_{-0.7}$  & $3.9\pm1.2^{c}$         \\
4U~1735--44     & $1.88\pm 0.48$  & 3.03 & $7\pm 2$             & $3.0^{c}$               \\
4U~1820--30     & $0.92 \pm 0.16$ & 1.52 & $2.7 ^{+0.4}_{-0.3}$ & $0.78\pm 0.03^{d}$      \\
GX~9+9         & $3.15\pm 0.21$  & 2.10 & $4.5^{+2.4}_{-1.2}$  &                         \\
Cygnus~X--1     & $6.23\pm 0.14$  & 8.10 & $5.35\pm 0.6$        & $6.6^{+0.8}_{-0.3}$ $^{e}$\\
			   &                 &      &                      & $5.4\pm 0.4^{f}$        \\
Cygnus~X--2     & $0.59 \pm 0.08$ & 2.20 & $2.3\pm 0.5$         & $1.9\pm0.5^{g}$         \\
Sco~X--1        & $1.10\pm 0.30$  & 1.47 &                      & $1.33\pm 0.02^{h}$      \\
			   &                 &      &                      & $1.1\pm 0.5^{i}$ \\
XTE~J1817--330 & $ 0.38\pm 0.20$ & 1.58 &                      &                         \\
\enddata
\tablenotetext{a}{\citet{dic90}.}
\tablenotetext{b}{\citet{jue04}.}
\tablenotetext{c}{\citet{cac09}.}
\tablenotetext{d}{\citet{cos12}.}
\tablenotetext{e}{\citet{mak08}.}
\tablenotetext{f}{\citet{han09}.}
\tablenotetext{g}{\citet{cos05}.}
\tablenotetext{h}{\citet{gar11}.}
\tablenotetext{i}{\citet{bra03}.}
\end{deluxetable}
%%%%%%%%%%%%%%%%%%%%%%%%%%%%%%%%%%%%%%%%%%%%%%%%%%%%%%%%%%%%%%%%%%%%%
Figure~\ref{fig10} shows for all sources a comparison of the oxygen, neon, and iron total column densities obtained from the {\tt ISMabs} broadband fit with those derived by \citet{jue04}. For an atomic element $X$, say, $N(X)=N{\left(X^{0}\right)}+N{ \left(X^{+}\right)}+N{ \left(X^{+2}\right)}$. The respective column densities by \citet{jue04} were obtained through a fit of eight galactic sources which includes those of the present report except for Sco~X--1 and XTE~J1817--330. The  average values for the oxygen column densities are $N({\rm O})=(12.96\pm 1.52)\times 10^{21}$~cm$^{-2}$ ({\tt ISMabs}) and $N({\rm O})=(24.00\pm 7.43)\times 10^{21}$~cm$^{-2}$ \citep{jue04}. In the case of neon we obtained an average value $N({\rm Ne})=(3.89\pm 0.57)\times 10^{21}$~cm$^{-2}$ ({\tt ISMabs}) compared to $N({\rm Ne})=(4.05\pm 1.06)\times 10^{21}$~cm$^{-2}$ \citep{jue04}. Lastly, for iron we estimated an average value of $N({\rm Fe})=(5.85\pm 1.12)\times 10^{21}$~cm$^{-2}$ ({\tt ISMabs}) compared to $N({\rm Fe})=(7.55\pm 1.76)\times 10^{21}$~cm$^{-2}$ by \citet{jue04}.

%%%%%%%%%%%%%%%%%%%%%%%%%%%%%%%%%%%%%%%%%%%%%%%%%%%%%%%%%%%%%%%%%%%%%
%%%%%%%%%%%%%%%%%%%%%%%%%%%%%%%%%%%%%%%%%%%%%%%%%%%%%%%%%%%%%%%%%%%%%

From such data we have estimated the $N({\rm O})/N({\rm Ne})$ and $N({\rm Fe})/N({\rm Ne})$ column-density ratios derived from the {\tt ISMabs} broadband fit. Neon is chosen as a reference since, as a noble gas, it is neither depleted to molecular nor to solid forms.  The present average ratio is $N({\rm O})/N({\rm Ne})=2.85\pm 0.37$, while that derived by \citet{jue06} is $N({\rm O})/N({\rm Ne})=3.7\pm 0.3$. Ratio differences may by due to systematic effects such as line saturation as quoted by \citet{gat13a} for the case of XTE~J1817--330.  The ratio derived by \citet{jue06} is higher than the present value; however, as discussed by these authors, the inclusion of black-hole systems (e.g. Cygnus~X--1) may modify results due to the presence of intrinsic emission or absorption features. We also obtain the average ratio $N({\rm Fe})/N({\rm Ne})=0.12\pm 0.02$ which is lower than solar, namely  $N({\rm Fe})/N({\rm Ne})=0.26$ \citep{gre98} and $N({\rm Fe})/N({\rm Ne})=0.48$ \citep{asp05}. This discrepancy may be due to either ISM Fe depletion or Ne overabundance. In fact, as previously discussed, we have found a Ne abundance greater than solar in the lines of sight toward 4U~1636--53, 4U~1735--44, GX~9+9, and XTE~J1817--330. The $N({\rm Fe})/N({\rm Ne})$ ratio by \citet{jue06} ($N({\rm Fe})/N({\rm Ne})=0.15\pm 0.01$), on the other hand, is in a fair agreement with {\tt ISMabs}. A further comparison of the iron and oxygen column densities finds the {\tt ISMabs} ratio $N({\rm Fe})/N({\rm O})=0.04\pm 0.01$ in close agreement with the solar ratio of \citet{gre98} (${\rm Fe/O}=0.04$) but somewhat lower than that by \citet{asp05} ($N({\rm Fe})/N({\rm O})=0.06$). Future analysis of the Fe K-edge absorption region would indeed help to constrain the ISM Fe abundance fractions in order to settle this issue.

The high-resolution X-ray spectra toward eight LMXBs have been analyzed. This study involved a broadband fit with {\tt ISMabs} to constrain the hydrogen column density, finding good agreement with previous work. Then, a simultaneous fit of the neon- and iron-edge regions was performed for each source, where the resulting column densities indicate the presence of a lowly ionized gas along the different lines of sight.

We intend to carry out in the near future a more extensive analysis of ISM photoabsorption using all the grating spectra from X-ray binaries available from the {\it XMM-Newton} and {\it Chandra} archives. Our target is to create a library of column densities and abundances for all the detectable atomic species in these spectra. We believe that once the atomic K edges are fully analyzed and modeled, the search for molecular and dust spectral features can be pursued in earnest, a topic of current interest in ISM astrophysics. Finally, the present {\tt ISMabs} model can be downloaded\footnote{http://heasarc.gsfc.nasa.gov/xanadu/xspec/models//ismabs.html} for use in X-ray data analysis packages such as {\sc xspec} \citep{arn96}, {\sc isis} \citep{hou00} and {\sc Sherpa} \citep{fre01}.

\acknowledgments
This work was supported by grant 15400673 of the Chandra Theory Program.

%%%%%%%%%%%%%%%%%%%%%%%%%%%%%%%%%%%%%%%%%%%%%%%%%%%%%%%%%%%%%%%%%%%%%
\begin{deluxetable*}{lllllllllc}
\tabletypesize{\tiny}
 \tablecaption{{\tt ISMabs} Broadband Simultaneous Fit Parameters \label{tab5}}
\tablewidth{0pt}
\tablehead{
\colhead{Source} &\multicolumn{8}{c}{{\tt ISMabs}} & \colhead{Statistics}\\
\cline{2-9}\\
\colhead{} & \colhead{$N({\rm H})$ } & \colhead{$N({\rm O~{\sc I}})$}& \colhead{$N({\rm O~{\sc II}})$}& \colhead{$N({\rm O~{\sc III}})$}& \colhead{$N({\rm Ne~{\sc I}})$}& \colhead{$N({\rm Ne~{\sc II}})$}& \colhead{$N({\rm Ne~{\sc III}})$}& \colhead{$N({\rm Fe})$}& \colhead{$\chi ^{2}$/dof  }
}
\startdata
4U~1636--53 &$4.16\pm 0.42 $ &$17.43 \pm 2.53  $ &$0.18 \pm 0.10  $&$ 0.58\pm 0.14 $& $4.00 \pm 0.67  $ &$0.78 \pm 0.37  $&$0.09\pm 0.01   $&$0.94\pm 0.19$& $ 3411/4313 $\\
4U~1735--44 &$1.88\pm 0.48  $ &$11.49  \pm 2.70  $ &$0.43 \pm  0.20 $&$0.49 \pm 0.11 $& $3.93  \pm 0.75  $ &$0.91  \pm 0.35  $&$0.04  \pm 0.02  $&$0.65  \pm 0.20  $& $ 1390/2293 $\\
4U~1820--30 &$0.92 \pm 0.16 $ &$6.99 \pm 0.93 $ &$0.45 \pm  0.32$&$0.22 \pm 0.15$& $1.25 \pm  0.26$ &$0.33 \pm 0.11  $&$0.05 \pm 0.02 $&$0.54 \pm 0.08 $& $6605 / 7943 $\\
GX~9+9 &$ 3.15\pm 0.21 $ &$12.30 \pm 1.12 $ &$0.35 \pm 0.28 $&$0.24 \pm 0.19$& $2.47 \pm 0.40 $ &$0.51 \pm 0.16   $&$0.05 \pm 0.03 $&$0.86 \pm 0.11 $& $ 3492/3277 $\\
Cygnus~X--1 &$6.23 \pm 0.14 $ &$23.38 \pm 0.72 $ &$0.16 \pm 0.12 $&$0.38 \pm 0.13$& $9.51 \pm 0.35 $ &$0.64 \pm 0.18  $&$0.14 \pm 0.05 $&$0.82 \pm 0.05 $& $7156 /4275 $\\
Cygnus~X--2 &$0.59 \pm 0.08 $ &$9.52 \pm 0.45 $ &$0.55 \pm 0.22 $&$0.20 \pm 0.07 $& $1.25 \pm 0.12 $ &$0.39 \pm 0.05  $&$0.04 \pm 0.01 $&$0.64 \pm 0.04 $& $ 8986/8941 $\\
Sco~X--1&$1.10 \pm 0.27 $ &$15.78 \pm 1.19 $ &$0.17 \pm 0.04 $&$0.00 \pm 0.00 $& (fixed) & (fixed)& (fixed)&$0.23\pm 0.08 $& $ 1520/1378 $\\
XTE~J1817--330&$0.38\pm 0.20 $ &$1.10 \pm 0.32 $ &$0.87 \pm 0.11 $&$0.46 \pm 0.07$& $1.56 \pm 0.10 $ &$0.66 \pm 0.04  $&$0.10 \pm 0.01 $&$0.46 \pm 0.23 $& $ 16335/9885 $\\
\enddata
\tablecomments{$N({\rm H})$ in  units of $10^{21}$~cm$^{-2}$. Metal column densities in units of $10^{17}$~cm$^{-2}$.}
\end{deluxetable*}
%%%%%%%%%%%%%%%%%%%%%%%%%%%%%%%%%%%%%%%%%%%%%%%%%%%%%%%%%%%%%%%%%%%%%

%%%%%%%%%%%%%%%%%%%%%%%%%%%%%%%%%%%%%%%%%%%%%%%%%%%%%%%%%%%%%%%%%%%%%%%%%%%%%%%

\appendix
\section{Role of ionization balance on abundances determination}\label{appendix}

%%%%%%%%%%%%%%%%%%%%%%%%%%%%%%%%%%%%%%%%%%%%%%%%%%%%%%%%%%%%%%%%%%%%%
\begin{figure*}[b]
  \epsscale{0.7}
  \plotone{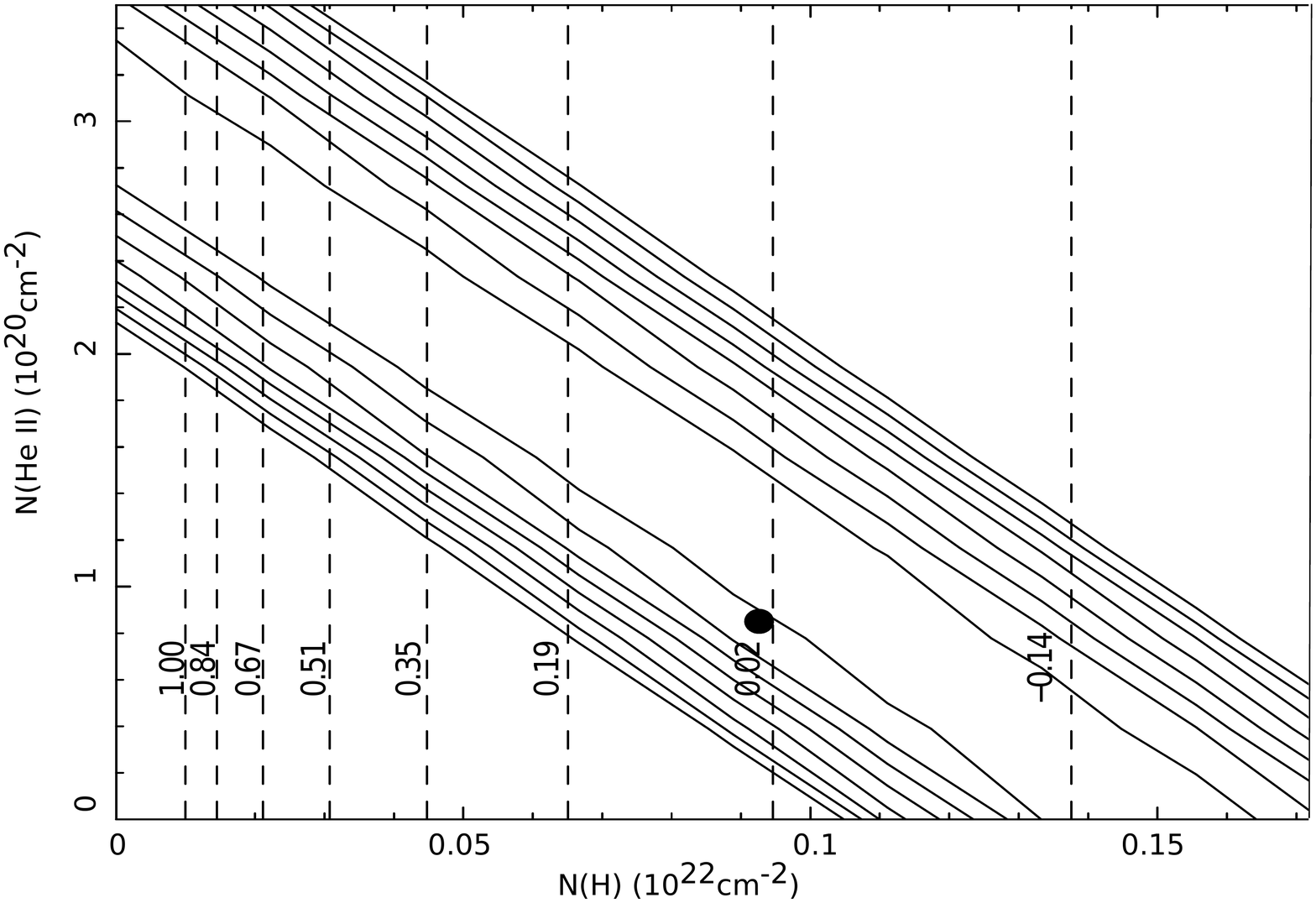}
  \caption{Constant $\chi^2$ contours for the broadband fits of the XTE~J1817--304 spectrum. The axes correspond to He~{\sc ii} and H column densities. Vertical dashed lines show the degeneracy between these two quantities since the neutral H column cannot be uniquely determined. The adopted {\tt warmabs} best-fit value is denoted with a black dot. \label{fig11}}
\end{figure*}
%%%%%%%%%%%%%%%%%%%%%%%%%%%%%%%%%%%%%%%%%%%%%%%%%%%%%%%%%%%%%%%%%%%%%

Here we discuss the differences in fitting spectra with the {\tt ISMabs} and {\tt warmabs} models, the latter as implemented in previous work \citep{gat14a, gat13a}. An important complication affecting our modeling concerns the smooth background photoelectric absorption which cannot be identified uniquely by spectral features.  This background is always present, and arises from hydrogen, helium, and valence-shell absorption by the same ions responsible for the edges and resonance absorption lines. Since all these mechanisms produce a smooth opacity, it is impossible to distinguish them uniquely.  If the gas is assumed to be in ionization balance, then the relative columns of neutral H, \ion{He}{1}, and \ion{He}{2} are determined; if not, it is important to be aware of the degeneracy between the smooth absorption produced by H, \ion{He}{1}, and \ion{He}{2}.

This degeneracy is illustrated graphically in Figure~\ref{fig11}, which shows the $\chi^2$ contours of the fit to the XTE~J1817--330 spectrum for various combinations of the H and \ion{He}{2} column densities.  The solid curves show contours of constant $\chi^2$ spaced at $\Delta \chi^2=100$. This clearly illustrates that the combinations of these variables lying along the diagonal are equivalent while those that fit the data have an almost exact inverse relationship. This obvious behavior must be understood and accounted for when deriving abundances relative to H or He in X-ray spectral fitting. In the results reported in the main body of this paper, which are based on {\tt ISMabs}, we have assumed that all the smooth opacity in the X-ray band comes from H + \ion{He}{1} and that the \ion{He}{2} column densities are negligible.

The {\tt warmabs} model includes the results of a photoionization equilibrium calculation, and therefore, the H and \ion{He}{2} column densities are tied together. The value taken from the {\tt warmabs} fit of the XTE~J1817--330 spectrum is shown with a dot in Figure~\ref{fig11}.  This corresponds to an ionization parameter of $\log\xi=-2.5$ for which the column density of neutral H  + \ion{He}{1} is 10.8 times that of \ion{He}{2}; i.e., the equilibrium assumption removes the degeneracy between the two variables.

A quantity of interest associated to the fitting of interstellar X-ray absorption spectra is the abundance of elements such as oxygen and neon relative to hydrogen. If the smooth background absorption in the X-ray band were solely due to neutral H, then the H and O or Ne column densities derived from the fits could be used to derive the relevant abundance ratios by simply taking column-density ratios. The degeneracy between \ion{He}{2} and H complicates this procedure since the total  H column density cannot be uniquely determined. This is shown graphically in Figure~\ref{fig11} with the vertical dashed curves which are curves of constant O abundance relative to H for the fit to the spectra of XTE~J1817--330.  The curves are labeled according to the conventional notation: $[{\rm O}]=0$ corresponds to the solar oxygen abundance \citep{gre98}, $[{\rm O}]=1$ to $10\times$ solar, etc.  This shows that, although acceptable fits can be obtained for any value of $N({\rm H})$ in the range up to $2\times 10^{21}$~cm$^{-2}$, the inferred abundance of O relative to H can take any value from 0.75 solar to 10 times solar.

Finally, we mention the unavoidable limitation that different numerical model implementations, even if they are based on the same atomic data, inevitably differ due to numerical details. For example, calculations in {\tt ISMabs} use cross sections tabulated on a grid that is finer than most instrumental grids, but which inevitably requires interpolation or smoothing to produce a model for {\sc xspec}. {\tt warmabs}, on the other hand, adopts such tabulations for some cross sections while for others analytic expressions are employed. Examples of the latter are the all-important cross sections for neutral O, Ne, and Mg of \citet{gor13}. Also, is important to note that the Fe cross-sections are different between both models, {\tt ISMabs} uses \citet{kor00} cross-section while {\tt warmabs} uses \citet{ver95} cross section. Small numerical differences in the cross section or opacity are exponentially amplified in the transmission coefficient which is most closely related to the observed spectrum. Thus, in spite of efforts to minimize numerical differences between codes, they are unavoidable at some level and it is useful to attempt to quantify them.

%%%%%%%%%%%%%%%%%%%%%%%%%%%%%%%%%%%%%%%%%%%%%%%%%%%%%%%%%%%%%%%%%%%%%
\begin{figure*}
  \epsscale{0.7}
  \plotone{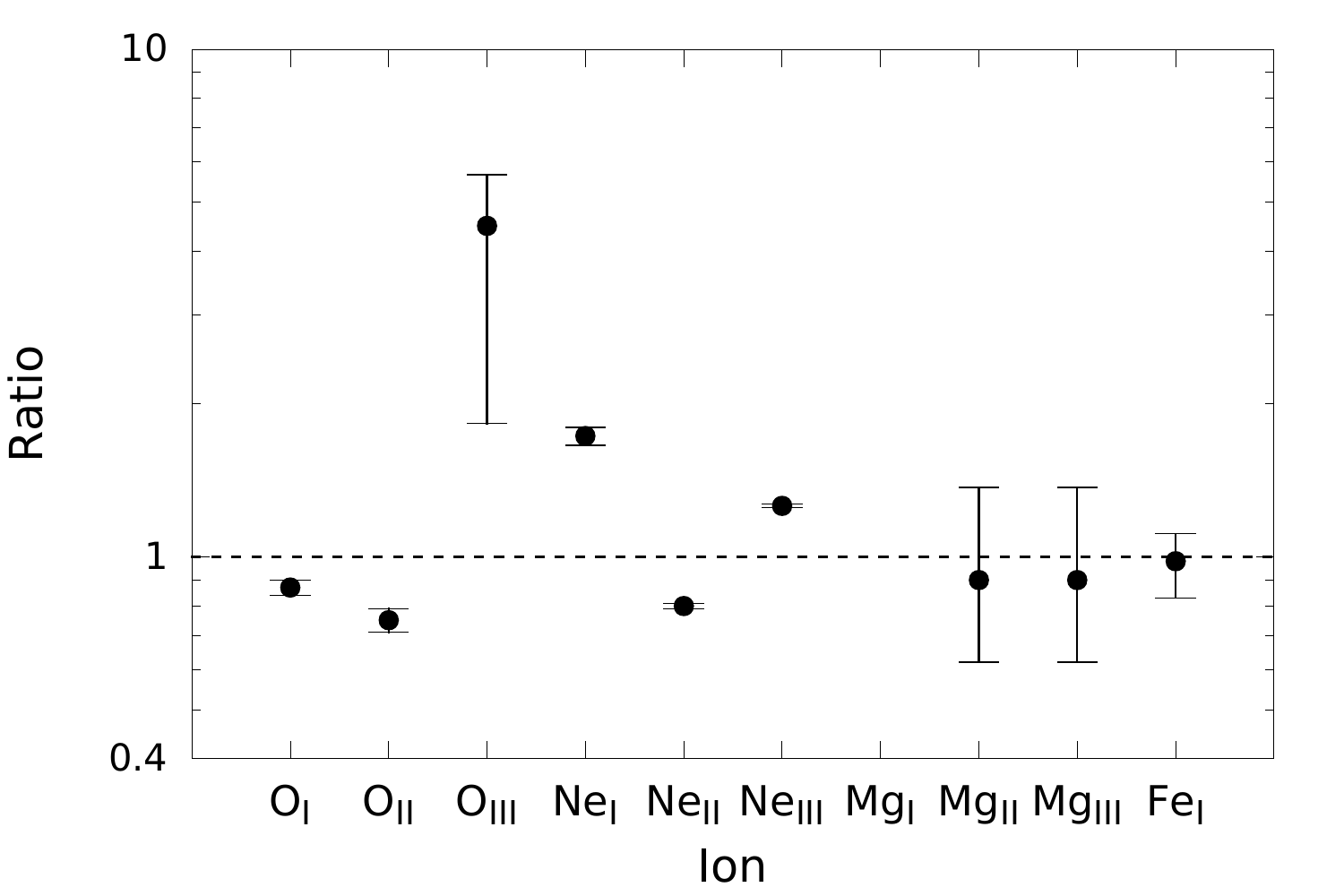}
  \caption{Ion column densities obtained from the fit to the XTE~J1817--304 spectrum. Each point corresponds to the {\tt ISMabs}/{\tt warmabs} ratio.\label{fig12}}
\end{figure*}
%%%%%%%%%%%%%%%%%%%%%%%%%%%%%%%%%%%%%%%%%%%%%%%%%%%%%%%%%%%%%%%%%%%%%

To summarize, we identify the three principal issues that affect the quantitative results of fitting to interstellar absorption and model comparisons: (i) the importance of ions and whether they are treated in equilibrium ratios or not; (ii) the ambiguity associated to the smooth background; and (iii) numerical implementation issues.  We now proceed to illustrate them with quantitative examples.

In Figure~\ref{fig12} we show a comparison of the ion column densities derived from fitting to XTE~J1817--330 with {\tt warmabs} and {\tt ISMabs}. In the case of {\tt warmabs}, we adopt two ionization components with $\log\xi=-5$ and $\log\xi=-2.5$, both with a column density of 10$^{21}$~cm$^{-2}$. The elemental abundances for the two components are allowed to vary freely in order to obtain the best fit. The four observations of this source are fitted simultaneously, and all the absorption parameters (i.e. the {\tt warmabs} ionization parameters, column densities, elemental abundances, turbulent velocities, and redshifts) are forced to be the same for all the different observations. The continuum is taken to be the sum of a black body and a power law, and the parameters of these components are allowed to differ between the various observations to reflect source variability.  The best fit has a $\chi^2/\nu=28325.47/15183=1.866$. {\tt warmabs} can output a list of ion column densities (even though these are not free parameters), and in Figure~\ref{fig12} we show a comparison of the {\tt warmabs} ion column densities with those from an {\tt ISMabs} fit. In the latter the H and \ion{He}{2} column densities are fixed at the values taken from the {\tt warmabs} fit. We take the column densities from all the ions from the {\tt warmabs} best fit and put them into {\tt ISMabs}.  This includes the H, \ion{He}{1}, and \ion{He}{2} column densities. Using the same continuum parameters we obtain $\chi^2/\nu=29876/15183=1.968$. This $\chi^2$ difference is likely to be numerical since all other parameters of the two fits are identical. We then allow the ion column densities in {\tt ISMabs} and the continuum parameters to vary freely in order to obtain the best fit. In doing so we do not allow the H, \ion{He}{1}, and \ion{He}{2} column densities to vary, and we preserve the restriction that the absorption model is the same for all observations. This procedure produces a fit with $\chi^2/\nu=27203/15153=1.795$.

%%%%%%%%%%%%%%%%%%%%%%%%%%%%%%%%%%%%%%%%%%%%%%%%%%%%%%%%%%%%%%%%%%%%%
\begin{figure*}
  \epsscale{0.7}
  \plotone{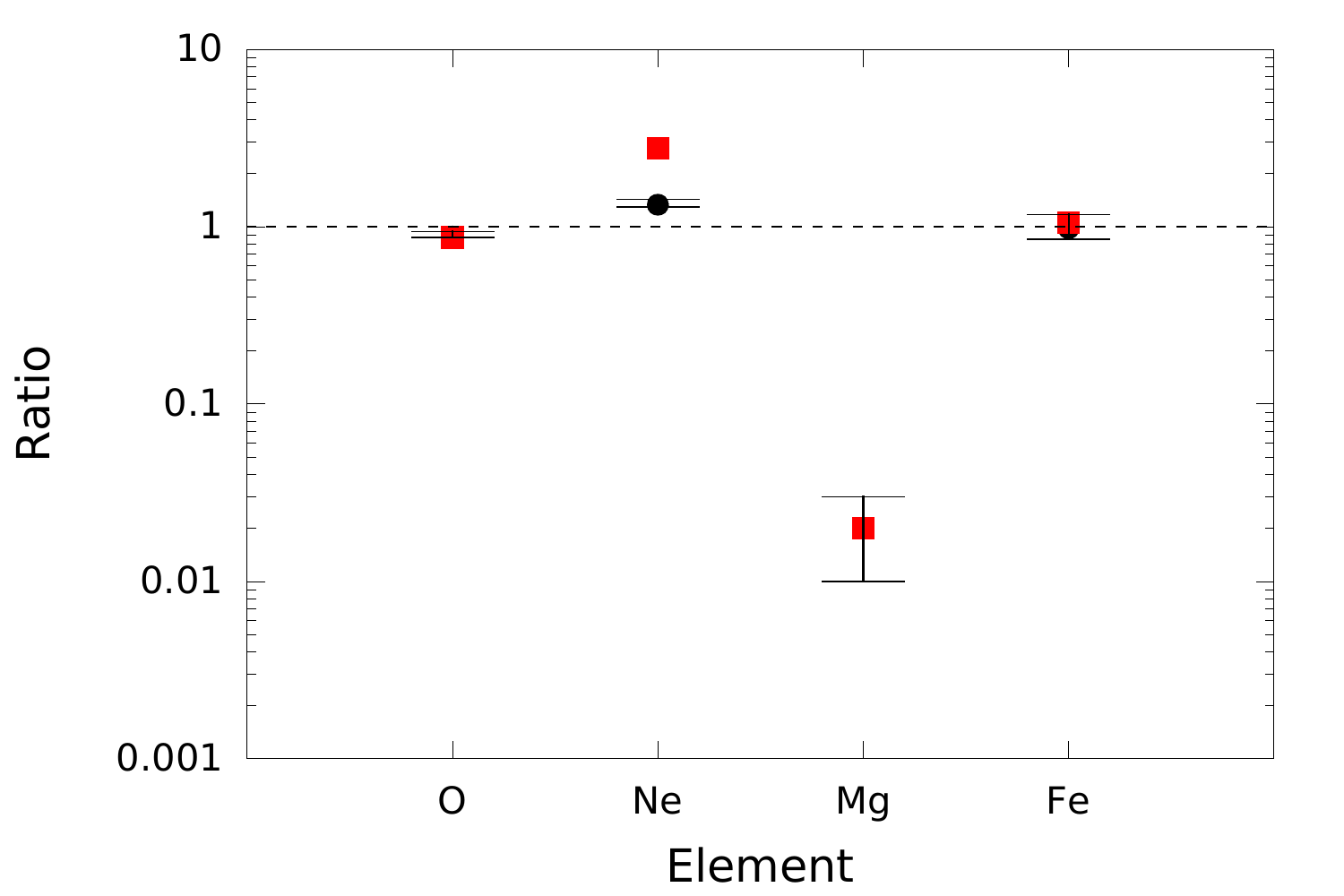}
  \caption{Abundances for various atomic elements obtained from the fit to XTE~J1817--304. Black circles correspond to the {\tt ISMabs}/{\tt warmabs} abundance ratios while the red squares indicate the ratios of {\tt ISMabs} to the solar abundances of \citet{and89}. \label{fig13}}
\end{figure*}
%%%%%%%%%%%%%%%%%%%%%%%%%%%%%%%%%%%%%%%%%%%%%%%%%%%%%%%%%%%%%%%%%%%%%

Figure~\ref{fig12} shows the ratio of the column densities derived from these two different methods. This represents both an astrophysical experiment on whether the ratios show photoionization equilibrium, and in some sense, a test of model accuracy (note that the Si and S column densities are negligible in both fits, so they are omitted). The two fits agree to within the error bars for most ions, a notable exception being \ion{O}{3} for which the {\tt warmabs} model predicts a significantly lower column density than {\tt ISMabs}. This indicates the existence of a component in the absorbing gas toward this source which is more highly ionized than the equilibrium gas that can actually fit most neutrals and singly ionized species. Oxygen shows such an effect more clearly than other elements due to its relatively high abundance.

A further step in exploring the effects of ionization equilibrium in absorption fitting is to sum the ion column densities derived from the fits with and without equilibrium (i.e. using {\tt warmabs} or {\tt ISMabs}) to get elemental abundances. Results are shown in Figure~\ref{fig13} where the ratio of elemental abundances are obtained by summing up the {\tt ISMabs} ion column densities to the elemental abundances obtained from the {\tt warmabs} column densities. The abundances shown here for {\tt ISMabs} make use of the H and \ion{He}{2} column densities derived by {\tt warmabs}. Also shown are the ratio of the {\tt ISMabs} abundances to the values in the compilation by \citet{gre98}. This shows that, in spite of the disagreement for \ion{O}{3},  both fitting methods yield abundances that are consistent with each other and with the compilation. However, the Ne abundance in this source is an exception; {\tt ISMabs} manifests a tendency to produce larger abundances since it is capable of accounting for ion absorption that is not tied up to the ionization equilibrium assumption.

%%%%%%%%%%%%%%%%%%%%%%%%%%%%%%%%%%%%%%%%%%%%%%%%%%%%%%%%%%%%%%%%%%%%%%%%%%%%%%%
\newpage
\bibliographystyle{apj}

%%%%%%%%%%%%%%%%%%%%%%%%%%%%%%%%%%%%%%%%%%%%%%%%%%%%%%%%%%%%%%%%%%%%%%%%%%%%%%%

\end{document}